\documentclass[12pt]{article}
\usepackage{amsmath}
\usepackage{amsfonts}
\usepackage[english]{babel}
\usepackage[ansinew]{inputenc}
\usepackage{graphicx}
\usepackage{graphpap}

\usepackage{psfrag}

\def\D{{\cal D}}
\def\defi{\stackrel{\mbox{\tiny \bf def}}{=}}
\def\vs{\vspace{5mm}}

\def\H{{\cal{H}}}
\def\Hxi{{\cal{H}}_{\xi}}
\def\gMm{g^{(n+1)}}

\def\eqH{\stackrel{\mbox{\tiny $\bm \H$}}{=}}

\def\eqSzero{\stackrel{\mbox{\tiny $\bm S_0$}}{=}}
\def\eqSone{\stackrel{\mbox{\tiny $\bm S_1$}}{=}}
\def\eqSf{\stackrel{\mbox{\tiny $\bm S[f]$}}{=}}
\def\Ric{\mbox{Ric}}
\def\Riem{\mbox{Riem}}
\def\Scal{\mbox{Scal}}\def\Ric{\mbox{Ric}}
\def\Riem{\mbox{Riem}}
\def\Scal{\mbox{Scal}}
\def\Ein{\mbox{Ein}}

\newcommand{\bm}[1]{\mbox{\boldmath $#1$}}

\newtheorem{definition}{Definition}
\newtheorem{lemma}{Lemma}

\newtheorem{theorem}{Theorem}
\newtheorem{corollary}{Corollary}
\newtheorem{Proposition}{Proposition}

\def\defi{\stackrel{\mbox{\tiny \bf def}}{=}}

\def\kellSzero{k}

\def\M{{\cal M}}
\def\Ein{{\mbox{Ein}}}
\def\gM{g^{(4)}}

\def\H{{\cal H}}

\def\N{{\cal N}}
\def\tr{\mbox{tr} \,}

\def\Journal#1#2#3#4#5#6{#1, ``#2'', {\em #3} {\bf #4}, #5 (#6).}

\def\JournalPrep#1#2#3{#1, ``#2'', #3.}

\def\JGP{\em J. Geom. Phys.}
\def\PLA{\em Phys. Lett. A}
\def\JDG{\em J. Diff. Geom.}
\def\CQG{\em Class. Quantum Grav.}

\def\PRD{{\em Phys. Rev.} \bm{D}}
\def\GRG{\em Gen. Rel. Grav.}

\def\JMP{\em J. Math. Phys.}

\def\CMP{\em Commun. Math. Phys.}

\def\PRL{\em Phys. Rev. Lett.}

\def\ATMP{\em Adv. Theor. Math. Phys.}

\def\IJMPD{\em Int. Jour. Mod. Phys. D}

\begin{document}

\title{Stability of MOTS in totally geodesic null horizons}
\author{Marc Mars \\
Facultad de Ciencias, Universidad de Salamanca,\\
Plaza de la Merced s/n, 37008 Salamanca, Spain \\
marc@usal.es}

\maketitle

\begin{abstract}
Closed sections of totally geodesic
null hypersurfaces are marginally
outer trapped surfaces (MOTS), for which a well-defined notion 
of stability exists. In this paper we obtain the explicit form for the
stability operator for such MOTS  and analyze in detail its properties
in the particular case of non-evolving horizons, which include both
isolated and Killing horizons. We link these stability properties with the
surface gravity of the horizon and/or to the existence of minimal 
sections. The results are used, in particular, to obtain
an area-angular momentum inequality for sections of axially
symmetric horizons in four spacetime dimensions, which helps
clarifying the relationship between two different approaches to
this inequality existing in the literature.
\end{abstract}


\section{Introduction}
\label{Intro}

Horizons play an important role in gravity theory and several different
types arise in many contexts. Particularly relevant are Killing
horizons, which are defined in spacetimes admitting a Killing vector
$\xi$ as null hypersurfaces 
$\Hxi$ where the Killing field $\xi$ is null, nowhere-zero
and tangent to $\Hxi$. An immediate consequence of this definition is that
any closed (i.e. compact and without boundary)
spacelike section of a Killing horizon is a marginally outer
trapped surface (MOTS for short), namely a closed codimension-two surface
with one of its null expansions identically vanishing.
MOTS are very interesting objects
both from a physical viewpoint, as quasi-local replacements
of black holes, and from a mathematical point of view, as objects
sharing several important properties with minimal hypersurfaces. In
particular, MOTS admit a sensible definition of stability, closely
related to the existence of outer trapped surfaces just inside
the MOTS. Since Killing horizons are foliated by MOTS, it is natural to
link the stability notion of its sections with the geometry of the Killing
horizon itself.

However, Killing horizons are unnecessarily restrictive in the sense that
their definition requires the existence of a special vector field
in the spacetime. The definition of Killing horizon can be relaxed
substantially by extracting the fundamental geometric properties
of $\Hxi$ which follow from the existence of a Killing vector and imposing
them directly on the null hypersurface.
This has lead to the
introduction of so-called non-expanding horizons,  and their
particularizations of weakly isolated horizons and isolated
horizons, which have been extensively studied in the literature.
{\it Non-expanding horizons} \cite{Haj1, Haj2, Ash1,
Ash1.5, Ash1.7, Ash2, GourgoulhonJaramillo2006, JaramilloPRD} are embedded null hypersurfaces $\H$
(usually with an additional topological restriction) with vanishing null
expansion and such that the Einstein tensor on $\H$ satisfies
an appropriate energy condition, which implies, among other things
that the   second fundamental form of $\H$ vanishes.
This, in turn, implies that
the null hypersurface admits a canonical connection ${\cal D}$
inherited from the spacetime.

To be more precise, define a null normal to $\H$
as a vector field $\ell$ which is null, nowhere zero and tangent to
$\H$. Null normals $\ell$ are obviously
defined
up to an arbitrary nowhere-zero rescaling. The energy
condition required in the definition of  non-expanding horizon is 
$\Ein(\ell, u ) |_{\H} \geq 0$ for any causal vector $u$ with the
same time-orientation than $\ell$  ($\Ein$ the Einstein tensor
of the spacetime).

%
%

Equivalent classes
 $[\ell]$ of null normals are defined by the
equivalence relation $\{ \ell^{\prime} \sim \ell$ if and
only if $\ell^{\prime} = c \ell\}$ with $c$ a nonzero constant. 
A {\it weakly isolated horizon} \cite{Ash2}
is a non-expanding horizon with a selected class of null normals
$[\ell]$ such that the commutator of the Lie derivative ${\mathcal L}_{\ell}$ 
and the covariant derivative ${\cal D}$ satisfies
$[ {\cal D}, {\mathcal L}_{\ell} ] \ell = 0$, for 
any $\ell \in [\ell]$. This property, in combination with the energy condition
above, implies that the surface
gravity along $\ell$, (i.e. the function $\kappa_{\ell}: \H \rightarrow
\mathbb{R}$ defined by $\nabla_{\ell} \ell = \kappa_{\ell} \ell$) is
constant. {\it Isolated horizons} \cite{Ash1} $\H$ are weakly isolated horizon
with the additional property that $[{\cal D}, {\mathcal L}_{\ell} ] = 0$,
for any $\ell \in [\ell]$.
Killing horizons $\H_{\xi}$ are isolated horizons whenever  
the class of null normals $[\ell]$ is chosen to be $[\xi]$ and
provided the
spacetime satisfies the energy condition mentioned before.
Non-expanding, weakly isolated and isolated horizons
were first introduced and studied in 
four spacetime dimensions, but 
all the definitions and most of the results carry over
to arbitrary spacetime dimension \cite{Lew1, Lew2}.

Although the energy condition imposed on non-expanding
horizons is physically reasonable, 
in geometric terms it is perhaps more natural to impose 
conditions directly on the geometry of $\H$ irrespectively of the 
energy-contents of the spacetime.
As already mentioned, the main consequence of the energy 
condition for non-expanding horizon is that the null
hypersurface is totally geodesic. It thus becomes of interest to analyze the geometry
of such hypersurfaces (this was in fact the approach taken in the seminal
paper by P. H\'{a}\'{\j}i\v{c}ek  \cite{Haj1}). In this context,
isolated horizons are naturally replaced by totally geodesic null
hypersurfaces with a selected null normal $\ell$ satisfying 
$[{\cal D}, {\mathcal L}_{\ell} ]=0$. Although  closely related 
to isolated horizons, this type of hypersurfaces is clearly more general. We
call them {\it non-evolving horizons} in this paper.

Similarly as for Killing horizons, any
closed  spacelike section of a totally geodesic null hypersurface is a MOTS.
It is therefore of interest to try and relate the stability properties of those
MOTS with the geometry of the null hypersurface. 
Isolated horizons admit several notions of extremality
\cite{BoothFairhurst2008}. One of them involves
the existence of trapped surfaces just inside the horizon. 
This type
of horizons are called subextremal, and they have played an important role
in the proof of area-angular momentum inequalities \cite{Hennig2008} 
in the case of axially symmetric horizons in four spacetime dimensions.
The subextremality condition of horizons is closely related to the stability
of the MOTS embedded in the horizon, so studying in detail the stability 
of these sections also clarifies under which conditions an isolated horizon
is subextremal or not.

Our aim in this paper is to perform a detailed study of the stability of 
MOTS in totally geodesic null hypersurfaces and, particularly, in 
non-evolving horizons. We obtain a simple and explicit form for the 
stability operator of MOTS lying in totally geodesic
null horizons (Proposition \ref{stability} and Corollary \ref{InTerms_of_z}).
As a
direct consequence we find (Corollary \ref{degenerate})
that non-evolving horizons
with vanishing surface gravity are always marginally stable.
We also prove  (Proposition \ref{marginal}) a result that
relates the stability properties of the sections with the sign of the surface
gravity
of the horizon, provided the MOTS is a marginally trapped surface.
This extends previous results obtained by Booth and Fairhurst 
\cite{BoothFairhurst2008} in the case
of spacetime dimension four and axially symmetric isolated horizons. We also 
study the dependence of the stability properties with the choice of
section in the horizon. We find that, in the constant surface gravity case, the
stability operator transforms nicely with the change of section and that
the stability properties are independent of the section. However, perhaps contrarily to
our intuition, the same is not true in the case of non-constant surface
gravity.
We give in Lemma \ref{counter}
an explicit example of a Killing horizon with non-constant surface
gravity where the stability depends on the choice of section.

One of our main results (Theorem \ref{minimality})  states that non-evolving
horizons with non-zero and constant surface gravity are marginally stable
if and only if they admit a section with vanishing total null expansion.
Under suitable additional conditions  (which are automatically satisfied
in static Killing horizons or in axially symmetric non-evolving horizons
of spherical topology) this section must, in fact, be minimal. An interesting
consequence of the results in this paper is that they allow 
us to improve our understanding of 
the relationship between the area-angular momentum
inequality proved by Hennig, Ansorg and Cederbaum \cite{Hennig2008}
in the context of stationary and axially symmetric four dimensional
spacetimes and the local proof obtained by Jaramillo, Reiris and Dain
\cite{JaramilloReirisDain} for stable MOTS. Some initial insight
on the relationship
between the two approaches has been recently obtained by
Jaramillo and Gabach-Cl\'ement \cite{MariaPepe} by showing that 
the key integral consequences of ``subextremality'' and
``strict stability'' employed respectively
in \cite{HennigCederbaumAnsorg2} and \cite{JaramilloReirisDain} 
(in the context of an area-angular momentum-charge inequality)
translate exactly into one another under a suitable
renaiming of functions. Recently a clarification
of the relationship between the inequality of Hennig {\it et al.} 
\cite{Hennig2008} and
another area-angular momentum inequality due
to Dain and Reiris \cite{DainReiris} and valid
for minimal surfaces lying in maximal
spacelike hypersurfaces in vacuum has also been obtained
\cite{ChruscielEckstein} by working in suitable coordinate systems.
Our approach here shows, in a purely geometric and coordinate
independent manner, how the framework for the general local inequality
for stable MOTS in \cite{JaramilloReirisDain} relates to the framework 
for the inequality in stationary and axially
symmetric black hole horizons \cite{Hennig2008}

The plan of the paper is as follows. In Section \ref{Notation}
we introduce our 
notation and basic definitions and recall the concept
of stability operator for MOTS. In Section \ref{Totally} we obtain the explicit
form of the  stability operator for sections of totally geodesic
null hypersurfaces. Section \ref{Non-evolving}
is devoted to analyzing the stability
properties of sections of non-evolving horizons. Finally in Section
\ref{Axially} we apply our results to the problem of area-angular momentum
inequalities for axially symmetric horizons in four spacetime dimensions.
In Theorem \ref{inequality} we find conditions, well-adapted to the horizon
setting, under which 
the area-angular momentum inequality holds. This result is then used
to  clarify the relationship between the results of Hennig {\it et
al.} and of Jaramillo {\it et al.} mentioned above. We finish the paper with 
a comment on the role played by the area-angular momentum inequality
in the proof of non-existence of two-black hole configurations
in equilibrium.

\section{Notation and basic results}
\label{Notation}

Throughout this paper, a spacetime $(M,\gMm)$ 
is an $(n+1)$-dimensional oriented manifold $M$ together
with a smooth metric $\gMm$ of Lorentzian signature. We assume
$(M,\gMm)$ to be time-oriented. All manifolds in this paper are connected and without boundary and all geometric objects are assumed to be smooth. 
Scalar product with a metric $h$ is denoted by $\langle \,\, , \,\, \rangle_{h}$
except for the spacetime metric, where we simply write
$\langle \,\, , \,\, \rangle$. The covariant derivative
of a metric $h$ is $\nabla^h$ and the corresponding Riemann tensor is denoted by
$\Riem_h$ (our sign conventions are $\Riem_h(X,Y) Z \defi (\nabla^h_X
\nabla^h_Y Z - \nabla^h_Y \nabla^h_X Z - \nabla^h_{[X,Y]} Z$). 
The Ricci, Einstein and 
curvature scalar of $\Riem_h$ are denoted by $\Ric_h$, $\Ein_h$ and $\Scal_h$ (with the sign
convention that the Ricci tensor and curvature scalar are positive for a standard sphere).
For the spacetime metric $\gMm$ we write $\nabla$, $\Riem$, $\Ric$,
$\Ein$ and $\Scal$. A spacetime is said to satisfy the {\bf dominant
energy condition} if $-\Ein(u,\cdot)$ is causal and future directed
for any future causal vector $u$. Note that this condition, when evaluated
on $\H$ with $u = \ell$ is precisely the energy condition imposed in the definition
of isolated horizon. Spacetime tensors carry Greek indices, which
are lowered and raised with $\gMm$.
Index components of the Riemann tensor are
written as $R^{\alpha}_{\,\,\beta\gamma\delta}$ and are defined by
$R^{\alpha}_{\,\,\beta\gamma\delta}  \defi  (\Riem(e_{\gamma},e_{\delta}) e_{\beta})^{\alpha}$.

In this paper we use the term ``spacelike surface'' to denote a closed (i.e. compact and without boundary)
spacelike, orientable, codimension-two embedded submanifold 
of $(M,\gMm)$. We will denote by $\Phi_S$ the embedding of $S$ into $M$ and we will
often identify $S$ and its image in $M$ under this embedding.
The induced metric in $S$ is denoted by $h$ and our convention for the second fundamental
form and mean curvature are
$\chi(X,Y) \defi - \left ( \nabla_X Y  \right )^{\bot}$ and $H \defi \tr_h \chi$,
where a vector $u \in T_p M, p \in M$, is decomposed as
$u = u^{\parallel} +u^{\bot}$ according to the direct sum decomposition $T_p M = T_p S \oplus N_p S$ of tangent and normal
spaces to $S$. The second
fundamental form along a normal $n$ is defined as $\chi_{n} \defi \langle \chi,n \rangle $.
Its trace defines the null expansion along $n$, $\theta_{n} \defi \tr_h \chi_{n}$.
The normal bundle of $S$, $NS = \bigcup_{p \in S} N_p S$  admits a global basis on null
vectors $\{\ell,k\}$ which we always take future-directed and partially normalized by $\langle \ell, k \rangle = -2$
(this leaves the usual boost freedom $\ell^{\prime} = F \ell$, $k^{\prime} = F^{-1} k$, where $F$ is a positive function on $S$).
Given a null basis $\{ \ell ,k \}$, the connection one-form of the normal bundle is denoted by ${\bm{s_{\ell}}}$ and defined as
\begin{eqnarray*}
\label{connection}
\bm{s_{\ell}} (X) \defi - \frac{1}{2} \langle k, \nabla_X \ell \rangle.
\end{eqnarray*} 

An important notion for spacelike surfaces is the first variation of its null expansions $\theta_{\ell}$ and $\theta_k$.
An explicit form for this variation has been obtained by several authors 
\cite{Newman1987, Hayward1994, CaiGalloway2001, GourgoulhonJaramillo2006, BoothFairhurst2007, AMS2008}.  
In most cases, the derivation uses implicitly
that the variation vector $\zeta$ is nowhere zero on $S$. This implies that, given any extension of $\zeta$ (which we may
assume to be  compactly supported without loss of generality)
to a neighbourhood of  $S$ and  the corresponding local one-parameter family of diffeormorphism $\varphi_{\tau}$, $\tau
\in I_{\epsilon} \defi (-\epsilon, \epsilon) $, we have the property (after restricting  $\epsilon$ if necessary)
that the map $\Psi_{\zeta}: S \times I_{\epsilon} \rightarrow
M$ defined by $\Psi_{\zeta} (p , \tau) \defi \varphi_{\tau} \circ \Phi_S(p)$ is an embedding. 
This assumption simplifies notably the calculation of the variation because one can work with Lie derivatives
applied to spacetime tensors (see, however, \cite{AMS2008} for a derivation which does not make
this implicit assumption). In this paper, we only need the variation formula when the variation vector $\zeta$ is nowhere zero. Let
$\ell^{(\zeta)}$ be a nowhere zero, future-directed
null vector on the hypersurface $\Sigma_{\zeta} \defi \Psi_{\zeta} (S \times I_{\epsilon} )$ with 
the property of being orthogonal to all surfaces $S_{\tau} \defi \varphi_{\tau} (S)$, 
$\tau \in (\epsilon,\epsilon)$. A superindex $(\zeta)$ is added because the vector field $\ell^{(\zeta)}$ obviously depends
on the variation vector $\zeta$. Note that, given $\zeta$, the field ${\ell}^{(\zeta)}$ is defined up to 
a rescaling $\ell^{(\zeta)} \rightarrow Q \ell^{(\zeta)}$, where $Q$ is a positive function on $\Sigma_{\zeta}$. Given $\zeta$
and $\ell^{(\zeta)}$, the first order variation of $\theta_{\ell}$ is defined as $\delta_{\zeta} \theta_{\ell} \defi \frac{d}{d\tau} 
\varphi^{\star} (\theta_{\ell_{\tau}}) |_{\tau=0}$, where $\theta_{\ell_{\tau}}$ is the null expansion of $S_{\tau}$ with respect to
the null normal $\ell^{(\zeta)} |_{S_{\tau}}$. This variation takes the explicit form
\begin{align}
\delta_{\zeta} \theta_{\ell} = & - \triangle_h \psi + 2 \bm{s_{\ell}} \left ( \nabla^h \psi \right )
+ \psi \left ( \mbox{div}_h \bm{s_{\ell}} - || {\bm s}_{\ell} ||^2_h + \frac{1}{2} \theta_{\ell} \theta_{k} +
\frac{1}{2} \Scal_h - \frac{1}{2} \Ein (\ell,k)
\right ) - \nonumber \\
& - \alpha \left ( \Ein(\ell,\ell) + || \chi_{\ell} ||^2_h 
\right ) + \kappa_{\zeta} \theta_{\ell},  
\label{variation1}
\end{align}
where $\alpha,\psi$ are defined by the decomposition $\zeta |_S = \alpha \ell - \frac{\psi}{2} k$ and
 $\kappa_{\zeta} \defi -
\frac{1}{2} \langle k , \nabla_{\zeta} \ell^{(\zeta)} \rangle |_S$. In this expression
$\triangle_h$ is the Laplacian on $(S,h)$, $\mbox{div}_h$ is the divergence with the metric $h$
and $\nabla^h \psi$ is the gradient of $\psi$. 
Interchanging $\{\ell,k \}$ in (\ref{variation1}) yields
\cite{BoothFairhurst2007}. 
\begin{align}
\delta_{\zeta} \theta_{k} = & - \triangle_h \psi
- 2 \bm{s_{\ell}} \left ( \nabla^h \psi \right )
+ \psi \left ( - \mbox{div}_h \bm{s_{\ell}} - || {\bm s}_{\ell} ||^2_h + \frac{1}{2} \theta_{\ell} \theta_{k} +
\frac{1}{2} \Scal_h - \frac{1}{2} \Ein (\ell,k)
\right ) - \nonumber \\
& - \alpha \left (  \Ein(k,k) + || \chi_{k} ||^2_h 
\right ) - \kappa_{\zeta} \theta_{k},  
\label{variation2}
\end{align}
where $\alpha, \psi$ are now defined by $\zeta |_S = \alpha k - \frac{\psi}{2}
 \ell$ (and both $\kappa_{\zeta}$ and $\bm{s_{\ell}}$ are the same as in 
(\ref{variation1}))

These expressions are well-suited for situations where one has good control on the geometric properties of $S$. However,
in some cases one has better knowledge of the properties of the variation vector $\zeta$. An appropriate
expression for the variation of the null expansion in this setting can be obtained in terms of the so-called,
{\bf deformation tensor} of $\zeta$, defined as $a^{\,\zeta\,} \equiv \mathcal{L}_{\zeta\,\,}\gMm$, where 
$\mathcal{L}$ denotes Lie derivative. The following proposition was proved in 
\cite{CarrascoMars2009} with the aim of analyzing the interplay between MOTS and symmetries
(given two 2-covariant tensors
$A, B$ on $(S,h)$, we define $A \cdot B = \tr_h (A \otimes B)$, where the
trace is taken with respect to the first and third indices).
\begin{Proposition}
\label{CM}
Let $S$ be a spacelike surface with embedding $\Phi_S$. With the notation introduced above,
the variation of  the null expansion $\theta_{\ell}$ along an arbitrary vector field
$\zeta$ takes the form 
\begin{align}
\delta_{\zeta}\,\theta_{k}
= (\mathcal{L}_{\zeta} {\bm k^{(\zeta)} }) (H) 
- \tr_h \left ( a^{\,\zeta, S} \cdot  \chi_{k} \right ) 
\left . + h^{\alpha\beta}   k^{\gamma} 
\left[ \frac{1}{2}\nabla_{\gamma}a^{\, \zeta \,}_{\alpha\beta}
- \nabla_{\alpha}a^{\, \zeta \,}_{\gamma\beta} \, \right]
\right|_{S},
\label{VariationDeformation1}
\end{align}
where ${\bm k^{(\zeta)}}$ is the one-form associated to $k^{(\zeta)}$, 
 $a^{\, \zeta, \, S} \defi \Phi^{\, \star}_S (a^{\, \zeta \,})$ and $h^{\alpha}_{\,\,\beta}$
is the projector tangent to $S$ (i.e. $h(v)= v^{\parallel}$).
\end{Proposition}
In this paper we will use this result in the following slightly modified form:
\begin{lemma}
\label{VariationDeformation2}
With the same notation as in Proposition \ref{CM}, we have
\begin{align*}
\delta_{\zeta}\,\theta_{k}
 = (\mathcal{L}_{\zeta} {\bm k^{(\zeta)} }) (H) 
- \tr_h \left . \left ( a^{\,\zeta, S} \cdot  \chi_{k} \right ) 
- h^{\alpha\beta}  k^{\gamma} \left ( \nabla_{\alpha} \nabla_{\beta}
\zeta_{\gamma} - R^{\mu}_{\,\,\,\,\alpha \beta \gamma} \zeta_{\mu} \right ) \right |_{S}. 
\end{align*}
\end{lemma}
{\it Proof.} The symmetry of $h^{\alpha\beta}$ allows us to write
\begin{eqnarray}
\left . h^{\alpha\beta}   k^{\gamma} 
\left[ \frac{1}{2}\nabla_{\gamma}a^{\, \zeta \,}_{\alpha\beta}
- \nabla_{\alpha}a^{\, \zeta \,}_{\gamma\beta} \, \right]
\right|_{S} = 
\left . \frac{1}{2} h^{\alpha\beta}   k^{\gamma} 
\left[ 
\nabla_{\gamma}a^{\, \zeta \,}_{\alpha\beta}
- \nabla_{\alpha}a^{\, \zeta \,}_{\gamma\beta} 
- \nabla_{\beta}a^{\, \zeta \,}_{\gamma\alpha} 
\, \right] \right|_{S} 
\label{sim}
\end{eqnarray}
Inserting $a^{\, \zeta}_{\alpha\beta} = \nabla_{\alpha} \zeta_{\beta} + \nabla_{\beta} \zeta_{\alpha}$ and using the Ricci
and first Bianchi identities it follows
\begin{eqnarray}
\nabla_{\gamma}a^{\, \zeta \,}_{\alpha\beta}
- \nabla_{\alpha}a^{\, \zeta \,}_{\gamma\beta} 
- \nabla_{\beta}a^{\, \zeta \,}_{\gamma\alpha}  = - 2 \left ( \nabla_{\alpha} \nabla_{\beta} \zeta_{\gamma}
- R^{\mu}_{\,\,\,\,\alpha\beta\gamma} \zeta_{\mu} \right ).
\label{Bian}
\end{eqnarray}
Combining (\ref{sim}) and (\ref{Bian}) into (\ref{VariationDeformation1}) proves the Lemma.
$\hfill \Box$

\vspace{5mm}

The variation formulae above are specially relevant for 
marginally outer trapped surfaces (MOTS), which 
are spacelike surfaces such that its mean curvature $H$ 
is everywhere parallel to one of the two null normals. Assuming that
$H \propto \ell$, a MOTS is defined by the property that $\theta_{\ell} =0$.
An important notion related to MOTS is the so-called stability
operator which is directly related to the first order variations of 
$\theta_{\ell}$ described above. More specifically, consider
section of the normal bundle which is nowhere
tangent to $\ell$. This section can be uniquely defined by a vector field
$v \in \mathfrak{X} (S)^{\bot}$ of the form
\begin{eqnarray*}
v = - \frac{1}{2} k + V \ell.
\end{eqnarray*}
where $V \in C^{\infty}(S,\mathbb{R})$.
It is useful to define the Hodge dual (see e.g. \cite{Bray2007})
\begin{eqnarray}
v^{\, \star} \defi \frac{1}{2} k +
V \ell,
\label{vstar}
\end{eqnarray}
which satisfies $ \langle v^{\, \star},v \rangle =0$ and
$ \langle v^{\, \star},v^{\, \star}\rangle = - \langle v,v \rangle $.
For any smooth function $\psi$ on $S$, the {\bf stability operator} \cite{AMS2008}
$L_{v}$ is the differential operator
defined by $L_{v} \psi \defi \delta_{\psi v} \theta_{\ell}$ .
Expression (\ref{variation1}) gives the explicit expression 
\begin{eqnarray}
L_v \psi= - \triangle_{h} \psi + 2 \bm{s_{\ell}} ( \nabla_{h} \psi) + \left ( 
\mbox{div}_{h} \bm{s}     -  ||\bm{s}||^2_{h} +  
\frac{\Scal_h}{2}  - \Ein(\ell, v^{\, \star}) -  V
|| \chi_{\ell} ||_h^2 \right ) \psi .
\label{stability1}
\end{eqnarray}
We will denote by $L_{-}$ the stability operator
along $- \frac{1}{2} k$. Expression (\ref{vstar}) implies 
\begin{eqnarray}
L_{v} = L_{-} - V ( \Ein(\ell,\ell) 
+ ||\chi_{\ell}||_h^2 ). \label{LvL-}
\end{eqnarray}
The stability operator, like any other second order elliptic operator
on a compact manifold, admits a principal eigenvalue, which is the
eigenvalue with lowest real part. This eigenvalue turns out to be real
and have a one-dimensional eigenspace (see the discussion in \cite{AMS2008}).
We will denote the
principal eigenvalue of $L_v$ by $\lambda_v$. The eigenspace of $\lambda_v$
is of the form $c \phi_v$ where $\phi_v>0$  and $c \in \mathbb{R}$.
According to the sign of the principal eigenvalue, a MOTS is called
strictly stable along $v$ if $\lambda_{v} >0$, marginally stable along $v$
if $\lambda_{v}=0$ and unstable along $v$ if $\lambda_v < 0$ \cite{AMS2008}.
A  MOTS is stable along $v$ iff $\lambda_v \geq 0$.

In the next section we consider totally geodesic null hypersurfaces,
which by construction
are foliated by MOTS, and we study the stability operator for those MOTS.

\section{Totally geodesic null hypersurfaces}
\label{Totally}
In this paper, a null hypersurface $\H$ is a codimension-one embedded
submanifold of $M$ with degenerate first fundamental form. 
As before, we usually identify $\H$ and its image in $M$.
Let $\ell$ be a future-directed null normal to $\H$, namely a nowhere-zero vector field 
on $\H$ which is null,  future-directed and tangent to $\H$. This vector field is
defined up to a rescaling $\ell^{\prime} = F \ell$ where $F : \H \rightarrow \mathbb{R}^+$. The one form $\bm{\ell}$ obtained by lowering its index defines a normal one-form
to $\H$. It is a standard property that the integral curves of $\ell$ are geodesic, which implies the
existence of  smooth function (the so-called {\bf surface gravity})
$\kappa_{\ell}:  \H \rightarrow \mathbb{R}$ satisfying
$\nabla_{\ell} \ell \eqH \kappa_{\ell} \ell$. Under a rescaling
${\ell}^{\prime} = F  \ell$, $\kappa_{\ell}$ transforms as $\kappa_{\ell^{\prime}} = F \kappa_{\ell} + \ell (F)$. 

The second fundamental form of $\H$ with respect to $\ell$ 
is defined
as $K_{\ell} (X,Y) \defi \langle Y , \nabla_X {\bm \ell} \rangle$, where $X,Y$ are vector fields tangent to
$\H$. Under a rescaling of $\ell$ we obviously have $K_{F \ell} = F K_{\ell}$. Null hypersurfaces
have the property that any spacelike surface embedded in $\H$ (i.e. such that the embedding
$\Phi_S : S \rightarrow M$ satisfies $\Phi_S (S) \subset \H$) has the property that $\ell |_S$ is automatically
a null normal to $S$. Moreover, given any pair $\{ X,Y\}$ of vector fields  tangent to $S$, the second
fundamental form along $\ell |_S$ of $S$ satisfies, at any $p \in S$,
$\chi_{\ell |_S} (X,Y) |_p  = K_{\ell} (X,Y ) |_p$. 

In this paper
we are interested in {\bf totally geodesic null hypersurfaces}, namely null hypersurfaces with
identically vanishing second fundamental form 
$K_{\ell}$ (the definition is obviously independent of the choice of null normal).
We assume further a topological condition, namely,

\vspace{3mm}

{\bf Topological condition:} The topology of $\H$ is 
$S \times \mathbb{R}$ where $S$ is a closed $(n-1)$-dimensional manifold. Furthermore, the null normal $\ell$
is tangent to the $\mathbb{R}$ factor. $\hfill$ $(\star)$

\vspace{3mm}

This topological condition implies $\H$ is a trivial bundle $(\H,S,\pi)$ where $\pi$ is projection of 
$S \times \mathbb{R}$ into the first factor. Moreover, it also implies
that spacelike surfaces embedded in $\H$ are automatically sections
of $(\H,S,\pi)$.
An immediate consequence of the vanishing of the second fundamental form
$K_{\ell}$ is that any embedded surface $S$ in $\H$ has vanishing second
fundamental form $\chi_{\ell}$
along the null normal $\ell |_S$. In particular, the null expansion $\theta_{\ell }$ vanishes identically,
so that $S$ is a MOTS. Another standard consequence is that
$\Ein(\ell,\ell) |_{\H} =0$, which follows immediately from the Raychaudhuri equation (or equivalently, from
(\ref{variation1}) with $\psi=0$).

Our aim in this section is to analyze the stability operator for 
sections of a totally geodesic
null hypersurface $\H$ satisfying the topological condition $(\star)$. Select
a null normal $\ell$ and choose  a section $S_0$. We will denote by
$k$  the unique (once $\ell$ is selected) 
future-directed null vector orthogonal to $S_0$ and
satisfying $\langle \ell,\kellSzero \rangle  |_{S_0}= -2$.  Since totally geodesic
null hypersurfaces satisfy $\Ein(\ell,\ell) \eqH 0$ and
$\chi_{\ell} \eqH 0$, it follows that the stability 
operator (\ref{LvL-}) of $S_0$
is independent of the section $v$ of the normal bundle. Thus,
the stability operator and the corresponding principal
eigenvalue are properties  of $S_0$ alone. We will denote them
by $L_{S_0}$ and $\lambda_{S_0}$ respectively. 


In order to find an expression
for $L_{S_0}$ which depends solely on geometric properties of $\H$
and of $S_0$, it is convenient to introduce the following tensor:
\begin{definition}
Let $\H$ be a totally geodesic null hypersurface and let $\ell$ be a null
normal of $\H$. Extend $\ell$ arbitrarily to a neighbourhood of $\H$. The
{\bf non-isolation tensor} is the tensor $\N^{\ell}$ defined at $p \in  \H$
by
\begin{align}
\N^{\ell} |_p : T_p \H \times T_p \H & \longrightarrow T_p \H \nonumber \\
X_p , Y_p & \longrightarrow \N^{\ell} |_p (X_p,Y_p) {}^{\gamma}
\defi  X_p^{\alpha} Y_p ^{\beta} \left ( \nabla_{\alpha} \nabla_{\beta} l^{\gamma} 
- R_{\mu\alpha\beta}^{\,\,\,\,\,\,\,\,\,\,\,\,\gamma} \ell^{\mu} \right ) |_p
\label{Tell}
\end{align}
\end{definition}

\vs

In order for this definition to make sense it is necessary to show
that $\N^{\ell}$ is independent of the extension of $\ell$ and that the right-hand side
of (\ref{Tell}) is tangent to $\H$.
\begin{lemma}
The tensor $\N^{\ell} |_p$ is well-defined.
\end{lemma}
{\it Proof.}
The fact that the second fundamental form $K^{\ell}$ vanishes implies that, for any pair
of vector fields $X,Y$ on $\H$, the vector field $\nabla_X Y$ is tangent to $\H$. This defines
a connection on $\H$ by ${\cal D}_X Y \defi \nabla_X Y$. Let us first show that $\N^{\ell} |_p (X,Y)$
is independent of the extension of $\ell$.  Extend arbitrarily $X|_p,Y|_p$ to tensor fields tangent to $\H$. We have
\begin{align*}
\N^{\ell} |_p (X_p,Y_p)  & = \left ( \nabla_X \nabla_Y \ell  - \nabla_{\nabla_X Y} \ell 
- \Riem(X,\ell) Y \right ) |_p \\ 
& = \left ( \D_X \D_Y \ell - \D_{\D_X Y} \ell -  \Riem(X,\ell) Y \right ) |_p, 
\end{align*}
which only depends on the values of $\ell |_{\H}$. In order to show that $\N^{\ell}|_p (X_p,Y_p)$ is tangent to $\H$ we calculate
the commutator of ${\mathcal L}_{\ell}$ and $D$. Let $X,Y,Z$ be arbitrary spacetime vector fields.
The following identity is well-known (and easy to prove)
\begin{eqnarray*}
\left ( {\cal L}_Z \left ( \nabla_X Y \right) - \nabla_{[Z, X]} Y - \nabla_X \left ( {\mathcal L}_{Z} Y \right ) 
\right )^{\gamma}
= X^{\alpha} Y^{\beta} \left ( \nabla_{\alpha} \nabla_{\beta} Z^{\gamma} - R_{\mu\alpha\beta}^{\,\,\,\,\,\,\,\,\,\,\,\,\gamma} Z^{\mu} 
\right )
\end{eqnarray*}
Thus, for $X,Y$ tangent to $\H$, 
\begin{align}
{\mathcal L}_{\ell} \left ( \D_X Y \right) - \D_{[\ell,X]} Y 
- \D_X \left ( {\mathcal L}_{\ell} Y \right )  
& = {\mathcal L}_{\ell} \left ( \nabla_X Y \right ) - \D_{[\ell,X]} Y  - \D_X \left ( {\mathcal L}_{\ell} Y \right )  
\nonumber
\\
& = \nabla_{[\ell,X]} Y + \nabla_X \left ( {\mathcal L}_{\ell} Y \right ) 
+ \N^{\ell} (X,Y) - \D_{[\ell,X]} Y - \D_X \left ( {\mathcal L}_{\ell} Y \right )  
\nonumber \\
& = \N^{\ell}(X,Y). \label{tangency}
\end{align}
Since the left-hand side is tangent to $\H$, so is the right-hand side. \hfill $\Box$

\vspace{3mm}

\noindent {\bf Remark.} The name {\bf non-isolation tensor} comes from the concept of isolated horizon.
As mentioned in the Introduction the definition of isolated horizon requires that 
the Lie derivative along $\ell$ and the covariant derivative $\D$ commute. Let, 
as before, $X,Y$ be arbitrary vector fields tangent to $\H$. 
The tensor $[{\mathcal L}_{\ell}, \D] Y$ is a one-covariant, one-contravariant tensor
which  acts on $X$ according to 
\begin{eqnarray*}
\left ( \left [ {\mathcal L}_{\ell}, \D  \right ] Y \right ) (X) = \left ( {\mathcal L}_{\ell} \left ( \D Y \right )
\right ) (X)  
- \D_X \left ( {\mathcal L}_{\ell} Y \right ) 
= {\mathcal L}_{\ell} \left ( \D_X Y \right ) 
- \D_{[\ell,X]} Y - \D_X \left ( {\mathcal L}_{\ell} Y \right ) =  \N^{\ell}(X,Y),
\end{eqnarray*}
where in the second equality we have used the Leibniz rule $({\mathcal L}_{\ell} T) (X) 
= {\mathcal L}_{\ell} ( T(X)) - T ( {\mathcal L}_{\ell} X)$
and the last equality follows from (\ref{tangency}). Thus, we conclude that ${\mathcal L}_{\ell}$ and $\D$ commute if and
only if $\N^{\ell}$ vanishes. This remark also shows that the non-isolation
tensor is a spacetime formulation of the tensor $-C^{c}_{ab}$ introduced in 
\cite{Ash2} (see eq. (4.3) there)

The following Proposition gives an explicit expression for the stability operator of $S_0$ in terms
of the properties of $S_0$ and of the totally geodesic horizon. This result extends a previous
result by Booth \& Fairhurst valid for isolated horizons \cite{BoothFairhurst2008}.

\begin{Proposition}[Stability operator of a section of a totally geodesic
null horizon]
\label{stability}
Let $\H$ be a totally geodesic null hypersurface satisfying the topological
condition $(*)$. Let $\ell$ be a null normal and $S_0$ a section of $\H$. 
Denote by $\kellSzero$ the unique future-directed null vector orthogonal to $S_0$ and
satisfying $\langle \ell, \kellSzero \rangle  |_{S_0}= -2$.  Then, the stability operator of $S_0$
reads
\begin{equation}
L_{S_0} \psi = - \triangle_h \psi + 2 {\bm s_{\ell}} \left ( \nabla^h \psi \right )
+ \psi \left ( 2 \mbox{div}_h \bm s_{\ell} - \frac{1}{2} \kappa_{\ell} 
\theta_{\kellSzero} + \frac{1}{2} \langle \kellSzero, \tr_h \N^{\ell} \rangle \right )
\label{stability2}
\end{equation}
\end{Proposition}

{\it Proof.} The fact that $\H$ is totally geodesic means precisely that the deformation tensor of $\ell$ vanishes
when acting on tangent vectors to $\H$. This implies, in particular, that $a^{\ell, S_0} \defi
\Phi_{S_0} (a^{\ell}) =0$.  Let $k^{(\ell)}$ be the vector field on $\H$ defined by the property that 
$k^{(\ell)}$ is null, satisfies $\langle \ell,k^{(\ell)} \rangle \eqH -2$ and
is orthogonal to $\varphi_{\tau} (S_0)$, where $\varphi_{\tau}$ is the local one-parameter group
of transformations generated by $\ell$. Under these conditions we can apply
Lemma \ref{VariationDeformation2} with $\zeta = \ell$.
Since the mean curvature of $S_0$ is $H = -  \frac{1}{2} \theta_{\kellSzero} \ell$, it follows
\begin{equation*}
 ( {\mathcal L}_{\ell}  {\bm k^{(\ell)}}  ) (H) = 
  - \frac{1}{2} \theta_{\kellSzero} 
( {\mathcal L}_{\ell}  {\bm k^{(\ell)}} )  (\ell) =
- \frac{1}{2} \theta_{\kellSzero} {\mathcal L}_{\ell}  (\langle \ell ,  k^{(\ell)} \rangle) = 0.
\end{equation*}
Using the definition of $\N^{\ell}$ we conclude, from Lemma \ref{VariationDeformation2},
\begin{equation}
\delta_{\ell} \theta_{\kellSzero} = -  \langle \kellSzero, \tr_{h} \N^{\ell} \rangle |_{S_0}.
\label{firstvar}
\end{equation}
Substituting $\alpha=0$ and $\psi=-2$ in (\ref{variation2}) yields
\begin{equation}
\delta_{\ell} \theta_{\kellSzero} =  
2 \mbox{div}_h \bm{s_{\ell}} + 2 || {\bm s}_{\ell} ||^2_h 
- \Scal_h + \Ein (\ell,\kellSzero) - \kappa_{\ell} \theta_{\kellSzero}.
\label{secondvar}
\end{equation}
Combining (\ref{firstvar}) and (\ref{secondvar}) it follows
\begin{eqnarray}
\Ein(\ell,\kellSzero ) \eqSzero \Scal_h - 2 \mbox{div}_h {\bm s_{\ell}}
- 2 ||{\bm s_{\ell}}||^2_{h} + \kappa_{\ell} \theta_{\kellSzero} -
\langle \kellSzero, \tr_h \N^{\ell} \rangle  \label{Einstein}
\end{eqnarray}
Substituting this into (\ref{variation1}) with $\alpha=0$ we find
\begin{equation*}
L_{S_0} (\psi) \defi \delta_{-\frac{\psi}{2} \kellSzero} \theta_{\ell} = 
- \triangle_h \psi + 2 {\bm s_{\ell}} \left ( \nabla^h \psi \right )
+ \psi \left ( 2 \mbox{div}_h \bm s_{\ell} - \frac{1}{2} \kappa_{\ell} 
\theta_{\kellSzero} + \frac{1}{2} \langle \kellSzero, \tr_h \N^{\ell} \rangle \right )
\end{equation*}
which proves the Proposition. \hfill  $\Box$

\vspace{5mm}

We note for later use the following expression obtained in (\ref{Einstein})
(c.f. expression (8.16) in \ref{GourgoulhonJaramillo2006}) 
\begin{corollary}
\label{CorEinxik}
The Einstein tensor of $(\M,\gMm)$ satisfies
\begin{eqnarray}
\Ein(\ell,k) \eqSzero \Scal_h 
-2 \mbox{div}_{h} \bm{s_{\ell}}  
- 2 ||\bm{s_{\ell}}||^2_{h} 
+ \kappa_{\ell} \theta_k - \langle k, \tr_h \N^{\ell} \rangle.
\label{Einxik}
\end{eqnarray}
\end{corollary}

\vs

Recall that on a compact Riemannian manifold $(S_0,h)$,
any one-form ${\bm \omega}$ admits a {\bf Hodge decomposition},
i.e. can be written as the sum of a gradient and
a divergence-free one-form, namely ${\bm \omega} = d f + {\bm \sigma}$
where $f \in C^{\infty} (S,\mathbb{R})$ and ${\bm \sigma}$
satisfies $\mbox{div}_{h} {\bm \sigma} = 0$. This decomposition is
unique except for an additive constant in $f$.

\begin{corollary}
\label{InTerms_of_z}
With the same notation and assumptions as in Proposition \ref{stability}.
Let  ${\bm s_{\ell}} = d Q_{\ell} + {\bm z}$ be the Hodge decomposition of ${\bm s_{\ell}}$
and define $u_{\ell} \defi e^{2Q_{\ell}}$. Then, the
stability operator of $S_0$ is
\begin{eqnarray}
L_{S_0} (\psi)  =
- \mbox{div}_{h} \left ( u_{\ell} \nabla_{h} \left ( \frac{\psi}{u_{\ell}} \right ) \right )
+ 2 \bm{z} ( \nabla_h \psi ) + \frac{1}{2}  \left ( 
\langle \kellSzero, tr_h \N^{\ell} \rangle 
- \kappa_{\ell} \theta_{\kellSzero} \right ) \psi,
\label{LS1}
\end{eqnarray}
\end{corollary}
{\it Proof.} Straightforward calculation. \hfill $\Box$

\vs

\noindent {\bf Remark.} Note that we have not added a subindex $\ell$ to the one-form $z$. This is because
this tensor is independent of the choice of $\ell$. Indeed, consider any other null normal
$\ell^{\prime} = F \ell$, with $F \in C^{\infty} (\H,\mathbb{R}^+)$ and define
$F_0 \defi F|_{S_0}$. From the definition of
connection one-form we have ${\bm s_{\ell^{\prime}} } = {\bm s_{\ell}} + d \, \ln F_0$. Consequently
${\bm z}$ does not depend on $\ell$ and $Q_{\ell^{\prime}} = Q_{\ell} + \ln F_0$   so that
$u_{\ell^{\prime}} = u_{\ell} F_0^2$.

\vs 

This expression for the stability operator of MOTS is the key to obtain the
stability results below. It is clear that knowledge of 
the non-isolation tensor is necessary in order to draw any conclusions. It
is reasonable to start with the simplest possible case, namely when this tensor vanishes
identically. This case is relevant because it includes all isolated horizons
and all Killing horizons (provided the topological condition $\star$ is satisfied).

\section{Stability of MOTS in non-evolving horizons}
\label{Non-evolving}

We start with the following  definition, which extends the usual notion of 
isolated horizon and has the advantage that it includes all
Killing horizons, irrespectively of whether the spacetime satisfies
suitable energy conditions or not.

\begin{definition}
Let $(M,\gMm)$ be a spacetime.
A {\bf non-evolving horizon} $(\H,\ell)$ 
is a totally geodesic null hypersurface
of $(M,\gMm)$ endowed with a future-directed null normal $\ell$
such that the non-isolation tensor $\N^{(\ell)}$ vanishes identically.
\end{definition}

This definition includes all Killing horizons because 
Killing vectors satisfy the well-known identity 
\begin{eqnarray*}
\nabla_{\alpha} \nabla_{\beta} \xi_{\gamma} = \xi_{\mu} 
R^{\mu}_{\,\,\,\alpha\beta\gamma}.
\end{eqnarray*}
Thus, with the canonical choice  $\ell = \xi |_{\Hxi}$ the hypersurface
$\Hxi$ satisfies $\N^{\ell}=0$ and $(\Hxi, \xi|_{\Hxi})$
is a non-evolving horizon.

Recall that a spacelike surface is future marginally trapped if its
mean curvature vector $H$ is future-directed and tangent everywhere
to one of its null normals. The following result
is an easy consequence of Corollary  \ref{InTerms_of_z}.
\begin{Proposition}
\label{marginal}
Let $(\H,\ell)$ be a non-evolving horizon satisfying the topological condition $(\star)$
and let $S_0$ be any section of
$\H$.
 Assume that  $S_0$ is future marginally trapped with non-identically
vanishing mean curvature $H$. Then
\begin{itemize}
\item If the surface gravity $\kappa_{\ell}$ is positive on $S_0$ then
$S_0$ is strictly stable.
\item If the surface gravity $\kappa_{\ell}$ vanishes on $S_0$
then $S_0$ is marginally stable.
\item If the surface gravity $\kappa_{\ell}$ is negative on $S_0$ then 
$S_0$ is unstable.
\end{itemize}
\end{Proposition}

\vs

\noindent {\bf Remark.} 
As already mentioned, isolated horizons always have
constant surface gravity $\kappa_{\ell}$. Thus, Proposition \ref{marginal} provides
a stability classification for future marginally trapped, non-minimal,
sections of isolated horizons. The sign of $\kappa_{\ell}$ determines the
stability character of $S_0$. In the particular case of axially symmetric
isolated horizons with topology $\mathbb{S}^2 \times \mathbb{R}$
and assuming that the mean curvature of $S_0$ is nowhere zero, this result
has been obtained by Booth and Fairhurst \cite{BoothFairhurst2008}. 
A generalization with the same symmetry and topology assumptions but assuming only $\theta_k \leq 0$ and negative somewhere
has been obtained by Jaramillo
\cite{JaramilloPrivate}.

\vs

{\it Proof.} Let $\lambda_{S_0}$ be the principal 
eigenvalue of the stability operator $L_{S_0}$. 
Let $\phi_0$ be principal eigenfunction $L_{S_0} (\phi_0) = \lambda_{S_0}
\phi_0$. Integrating this equation on $S_0$ we get, from (\ref{LS1}) with
$\N^{\ell}=0$,
\begin{equation}
\lambda_{S_0} \int_{S_0} \phi_0 \bm{\eta_{S_0}} = - \frac{1}{2} \int_{S_0}
\kappa_{\ell} \theta_{\kellSzero} \phi_0 \bm{\eta_{S_0}},
\label{integrate}
\end{equation}
where ${\bm \eta_{S_0}}$ is the volume form of $(S,h)$ and
we have used the fact that ${\bm{z}} (\nabla^h \psi)$ is a divergence
for any $\psi$ and hence integrates to zero. Since $\phi_0$
has constant sign, the claims of the Proposition
follow directly from  (\ref{integrate}) after using the fact that under the 
conditions of the Proposition $\theta_k |_{S_0} \leq 0$ and not identically
zero. \hfill $\Box$

\vs
In fact, in the degenerate case (i.e. when the surface gravity vanishes)
the argument proves a stronger statement.

\begin{corollary}
\label{degenerate}
Let $(\H,\ell)$ be a non-evolving horizon and let $S_0$ be any section of
$\H$. If $\kappa_{\ell}$ vanishes on $S_0$ then $S_0$ is marginally stable.
\end{corollary}

\vs

Having derived a general expression of the stability operator
of sections of a totally geodesic null hypersurface, it is natural to ask how does
this operator depend on the section. 
In order to determine the dependence on the section, we need to compare two arbitrary sections on $\H$.
Let us therefore fix a section $S_0$ on $\H$ and define a function
$\tau$ in $\H$ by $\ell (\tau)=1$ and $\tau |_{S_0} =0$. $\tau$ is
a coordinate along the $\mathbb{R}$ factor in $\H$. Any other section $S[f]$ 
can then be defined as a graph over $S_0$, $f: S_0 \rightarrow \mathbb{R}$. Let
$\pi_{f} : S[f] \rightarrow S_0$ be the natural projection along
orbits of $\ell$. $\pi_f$ is a diffemorphism and in fact an
isometry between these two spaces with their respective induced metrics.
Since the derivation holds independently of whether $\N^{\ell}$ vanishes or not, 
we state the result for general totally geodesics null hypersurfaces.


In order to determine the behaviour of the stability operator, we
need to relate the connection one-forms $\bm{s_{\ell}}$ of $S_0$ and $\bm{s_{\ell}[f]}$ of 
$S[f]$ and the null expansions $\theta_k$ of $S_0$ and 
$\theta_{k}[f]$ of $S[f]$. A related result written in terms of the
induced connection of the null hypersurface was obtained in \cite{Ash2} (see
also \cite{GourgoulhonJaramillo2006}) 
\begin{lemma}
\label{section}
Let $\H$ be a totally geodesic null hypersurface satisfying the topological condition
$(\star)$ and let $\ell$ a null normal to $\H$. Fix a section $S_0$ and
define $S[f]$ and $\pi_f$ as before.  Let $k_f$ be the
null normal to $S[f]$ satisfying $ \langle \ell  , k_f \rangle \eqSf -2$.
Denote by $\bm{s_{\ell}[f]}$ the normal connection of
$S[f]$, by $\chi^{k}[f]$ the second
fundamental form of $S[f]$ along $k_f$ and by $\theta_k[f]$ its trace. Then
\begin{eqnarray}
\bm{s_{\ell}[f]} & = &  \pi_f^{\star} \left ( \bm{s_{\ell}} + \kappa_{\ell} d f \right ), 
\label{first} \\
\chi^{k}[f]  & = &  \pi_f^{\star} \left (
\chi^{k} + 2 \mbox{Hess}_h f + 2 \kappa_{\ell} df \otimes df 
+ 2 df \otimes {\bm{s_{\ell}}} + 2 {\bm{s_{\ell}}} \otimes df \right )
\label{second}\\
\theta_k [f] & = & \left ( \theta_k
+ 2 \triangle_h f + 2 \kappa_{\ell} || \nabla^h f ||^2_h  + 4 \bm{s_{\ell}} \left ( \nabla^h f  \right ) 
\right ) \circ
\pi_f \label{third}
\end{eqnarray}
where  $\bm{s_{\ell}}$, $\chi^k$ and $\theta_k$ are, respectively,
$\bm{s_{\ell}}[f=0]$, $\chi^k[f=0]$ and $\theta_k[f=0]$ and
$\mbox{Hess}_h f$ is the Hessian of $f$ with the metric $h$.
\end{lemma}

{\it Proof}. As before, let $\varphi_{\tau}$ be the local one-parameter group of diffeomorphisms 
of $\H$ generated by $\ell$ and $S_{\tau} = \varphi_{\tau} (S_0)$. Let $k$ be the unique vector
field on $\H$ with the properties of being null, orthogonal to $S_{\tau}$ and 
satisfying $\langle k,  \ell \rangle = -2$. 
Given any vector field $X$ defined on  $S_0$ and tangent
to $\H$, we consistently 
extend it to $\H$ by solving ${\mathcal L}_{\ell} X =0$ with this initial
data. This vector field is obviously tangent to $S_{\tau}$.
We will refer to any such vector field as a {\it vector field
tangent to $S_0$} although it is defined everywhere on $\H$.
Any scalar function $w \in C^{\infty}(S_0,\mathbb{R})$  is also 
extended  to $\H$ by solving ${\mathcal L}_{\ell} w =0$ with this initial data.

With this notation it is  immediate to check that 
$X_f \defi ( \pi_f^{-1} )_{\star} (X)  = X + X(f) \ell$. 
For the null vector $k_f$ we note that $\langle  k_f, \ell \rangle \eqSf -2$
implies $k_f \eqSf k + A \ell + Z$ for some
function $A$ and some vector field $Z$ tangent to $S_0$.
Multiplying by $X_f$ we get
\begin{eqnarray*}
0 = \langle k_f, X_f \rangle = - 2 X(f) + \langle Z, X \rangle_h, 
\end{eqnarray*}
which implies $Z \eqH 2 \nabla^h f$. The condition $\langle k_f,k_f
\rangle \eqSf 0$ fixes
$A \eqH ||\nabla^h f||^2_h$. Thus,
\begin{eqnarray}
k_f \eqSf k + ||\nabla^h f||^2_h \ell + 2 \nabla^h f.
\label{Decompositionkf}
\end{eqnarray}
Let us start with the calculation of $\bm{s_{\ell}[f]}$. From its definition,
\begin{eqnarray}
\bm{s_{\ell}[f]} (X_f) & \defi &  - \left . \frac{1}{2} \langle k_f, \nabla_{X_f} \ell 
 \rangle \right |_{S[f]}  \nonumber \\
& = &  -\frac{1}{2} \langle  k + ||\nabla^h f ||^2_h
\ell + 2 \nabla^h f , \nabla_{X} \ell + X(f) \kappa_{\ell} \ell \rangle \nonumber \\
& = & {\bm{s_{\ell}}} (X) + \kappa_{\ell} X (f), \label{sf} 
\end{eqnarray}
where in the last equality we have used the fact 
that $\ell$ is orthogonal to any  vector tangent to $\H$ and 
\begin{eqnarray}
\langle  \nabla^h f, \nabla_{X} \ell \rangle = 0, \label{vanish}
\end{eqnarray}
which is a consequence of the vanishing of second fundamental form $K^{\ell}$ of $\H$.
Now, since $\pi_f^{\star} ({\bm{s_{\ell}}}) (X_f) = {\bm{s_{\ell}}}  [ (\pi_f)_{\star}
(X_f)] = {\bm s_{\ell}} [ (\pi_f)_{\star} \circ (\pi^{-1}_f )_{\star} (X)] =
{\bm{s_{\ell}}} (X)$, (\ref{sf}) can be written as
\begin{eqnarray*}
\bm{s_{\ell}[f]} (X_f) = \pi_f^{\star} \left (\bm{s_{\ell}} + \kappa_{\ell} df \right ) (X_f)
\end{eqnarray*}
which proves (\ref{first}). In order to show (\ref{second}) we need to evaluate the second fundamental
form of  $S[f]$ along $k_f$. Let $X,Y$ be a pair of vector fields
tangent to $S_0$ and $X_f,Y_f$ the corresponding fields tangent to $S[f]$. Using
${\cal L}_{\ell} Y = 0$
(in the form $\nabla_{\ell} Y = \nabla_Y \ell $), it follows by straightforward calculation
\begin{eqnarray*}
\nabla_{X_f} Y_f \eqSf \nabla_X Y + \left [ X (Y(f)) + \kappa_{\ell}  X(f) Y(f) \right ] \ell + X(f) \nabla_Y \ell
+ Y(f) \nabla_X \ell. 
\end{eqnarray*}
Multiplying this expression with $k_f$, as given in (\ref{Decompositionkf}), and using (\ref{vanish}) we find
\begin{align}
\chi^{k}[f] \left ( X_f, Y_f \right )  \defi & - \left . 
\langle  k_f, \nabla_{X_f} Y_f \rangle  \right |_{S[f]} \nonumber \\
= & \chi^{k} \left ( X , Y \right ) + 2 \left [
 X (Y (f)) - \langle \nabla^h f, \nabla_X Y \rangle  + \kappa_{\ell} X(f) Y(f)  \right ]  \nonumber \\
& 
+ 2 \left [ X(f) {\bm{s_{\ell}}} (Y) + Y(f) {\bm s_{\ell}}(X) \right ].
\label{SFF}
\end{align}
Now, by definition of Hessian, $\mbox{Hess}_h f (X,Y) = 
X(Y(f)) - \langle \nabla^h f, \nabla_X Y \rangle$, and (\ref{SFF}) becomes
\begin{eqnarray*}
\chi^{k}[f] (X_f,Y_f) & = & 
\chi^{k} (X,Y) + 2  \mbox{Hess}_h f \left ( X,Y \right ) + 2 \kappa_{\ell}
X(f) Y(f) + 2 \left [ X(f) {\bm{s_{\ell}}} (Y) + Y(f) \bm{s_{\ell}} (X) \right ]  \\
& = & \pi_f^{\star} \left (
\chi^{k} + 2 \mbox{Hess}_h f + 2 \kappa_{\ell} df \otimes df 
+ 2 df \otimes {\bm{s_{\ell}}} + 2 {\bm{s_{\ell}}} \otimes df \right ) (X_f, Y_f).
\end{eqnarray*}
This establishes (\ref{second}).  Taking the trace with the metric of $S[f]$
and using that $\pi_f$ is an isometry gives (\ref{third}).  \hfill $\Box$

\vs


\vs 

With this transformation lemma at hand, we can relate the stability
operators of $S_0$ and of $S[f]$ in the particular case of constant surface gravity.
As discussed above, this is the physically most interesting situation
since it holds for any spacetime satisfying the dominant energy condition.

\begin{Proposition}[Dependence of the stability operator on the section]
\label{Dependence_section}
Let $(\H,\ell)$ be a non-evolving horizon satisfying the topological condition $(\star)$
and assume that the surface gravity
$\kappa_{\ell}$ is constant. Then the stability operator of
$S[f]$ is related to the stability 
operator of $S_0$ by
\begin{eqnarray*}
L_{S[f]} \left ( \psi \circ \pi_f \right ) = e^{\kappa_{\ell} f} L_{S_0} \left (
e^{-\kappa_{\ell} f} \psi 
\right ) \circ \pi_f, \quad \quad \forall \psi \in C^2(S_0,\mathbb{R}).
\end{eqnarray*}
\end{Proposition}

{\it Proof.}  The equality is tensorial, so it suffices to work in a 
suitable coordinate system. Select any point
$p \in S_0$ and a coordinate system $\{U_p,y^A \}$  near $p$ ($A,B = 1,\cdots, n-1$). 
Then $\{ \pi_f^{-1} (U_p), y^A \circ \pi_f  \}$ is a coordinate system
near $\pi_f^{-1}(p)$. In these coordinates, $\pi_f$ is the identity map
and we can simply write
\begin{eqnarray*}
{\bm s_{\ell}[f]}_A & = &  \bm{s_{\ell}}_A + \kappa_{\ell} \nabla^h_A f, \\
\theta_k[f] & = &  \theta_k + 2 \triangle_h f + 2 \kappa_{\xi} \nabla^h_A f \nabla^{h\,A} f + 4 {s_{\ell}}^A \nabla^{h}_A f.
\end{eqnarray*}
Applying (\ref{stability2}) to the function $w \psi$ and expanding derivatives of products it follows, assuming $w>0$ everywhere,
\begin{equation}
\frac{1}{w}  L_{S_0} \left ( w \psi \right ) = - \triangle_h \psi + 2 
\left ( s_{\ell}^A - \frac{\nabla^{h \,A} w}{w} \right ) \nabla^h_A \psi + \left ( 2 
\nabla^h_A s_{\ell}^A - \frac{1}{2} \kappa_{\ell} \theta_k + 2 s^{A}_{\ell}  \frac{\nabla^h_A w}{w}
- \frac{1}{w} \triangle_h w \right ) \psi. \label{LS02}
\end{equation}
On the other hand, Lemma \ref{section} and (\ref{stability2})
imply that the stability operator for $S[f]$ reads, in these coordinates
(recall that $\pi_f$ is an isometry)
\begin{align}
L_{S[f]} (\psi) &=  - \triangle_h \psi + 2 s_{\ell}[f]^A \nabla^h_A \psi
+ \left ( 2 \nabla^h_A s_{\ell}[f]^A - \frac{1}{2}\kappa_\ell \theta_k[f] \right ) \psi \nonumber \\
& =  - \triangle_h \psi + 2 \left ( s_{\ell}^A + \kappa_{\ell} \nabla^{h \,A} f \right )
\nabla^h_A \psi + \left ( \frac{}{} 2 \nabla^h_{A} s_{\ell}^A + \kappa_{\ell} \triangle_h f - \right . \nonumber \\
&  \left .  \frac{1}{2}
\kappa_{\ell} \left ( \theta_k + 2 \kappa_{\ell} \nabla^h_A f \nabla^{h\,A} f
+ 4 s_{\ell}^A \nabla^h_A f \right ) \right ) \psi, \label{LSf2}
\end{align}
where we have used that $\kappa_\ell$ is constant. We want to find $w$ such that 
the right hand sides of (\ref{LS02}) and of (\ref{LSf2}) are the same. The term in $\nabla^h_A \psi$ 
forces $w = e^{-\kappa_{\ell} f}$ up to an irrelevant multiplicative constant. Inserting this expression in the zero
order term of (\ref{LS02}) we get the zero order term of (\ref{LSf2}),
which proves the claim. $\hfill \Box$

\vs

\begin{corollary}
\label{constancy}
Let $(\H,\ell)$ be a non-evolving horizon satisfying the topological condition $(\star)$
with constant surface gravity. Then
the principal eigenvalue is independent of the section.
\end{corollary}

{\it Proof. } Let $\phi_0$ be 
the principal eigenfunction and $\lambda_{S_0}$ the principal eigenvalue of $S_0$.
Define $\phi_{f} \equiv (e^{\kappa_{\ell}} \phi_0) \circ \pi_{f}$. Then
\begin{eqnarray*}
L_{S[f]} (\phi_{f} ) = L_{S[f]} \left ( (e^{\kappa_{\ell}} \phi_0 )\circ
\pi_f \right ) = (e^{\kappa_{\ell} f} L_{S_0} (\phi_0)) \circ \pi_{f} =
\lambda_{S_0} (e^{\kappa_{\ell} f} \phi_0) \circ \pi_f = \lambda_{S_0} \phi_{f}
\end{eqnarray*}
Thus, $\phi_{f}$ is an eigenfunction of $L_{S[f]}$ which must be the
principal eigenfunction because it does not change sign. $\hfill \Box$

\vs

This corollary states that, whenever the 
surface gravity is constant, the stability eigenvalue 
is a property of the non-evolving horizon  $(\H,\ell)$ instead
of a property of any particular section. In the constant surface gravity case,
we will denote this eigenvalue by $\lambda_{\H}$ and we will say that the horizon
is stable, marginally stable or unstable depending on the sign of $\lambda_{\H}$.

Proposition \ref{Dependence_section} and Corollary \ref{constancy} give a fully satisfactory answer to the issue of
how does stability depend on the section in the constant surface
gravity case. It is quite natural to ask whether the independence of the eigenvalue on the section
also holds in the case of non-constant surface gravity. The following
lemma shows that Corollary \ref{constancy} is not true when $\kappa_{\ell}$ is
not a constant (which, recall, can only happen
if the dominant energy condition does not hold).

\begin{lemma}
\label{counter}
There exist Killing horizons with non-constant surface gravity for which
the principal eigenvalue depends on the section.
\end{lemma}

{\it Proof.} It suffices to give an explicit example.
Consider the four-dimensional spacetime $\mathbb{R}^2 \times \mathbb{S}^2$
with metric
\begin{eqnarray*}
ds^2 = - 2 x W  dt^2 + 2 dt dx  + \gamma,
\end{eqnarray*}
where $\gamma$ is the standard metric on the sphere and $W : \mathbb{S}^2
\rightarrow \mathbb{R}$ is a smooth function which we take to be constant along 
one of the Killing vectors of $(\mathbb{S}^2,\gamma)$.
It is clear that $\xi = \partial_t$ is a Killing vector and $\Hxi 
\defi \{x=0\}$ is a Killing horizon of topology
$\mathbb{R} \times \mathbb{S}^2$. An immediate calculation shows that 
the surface gravity of $\Hxi$ is $\kappa_{\xi} = W$ and that 
the connection one-form $\bm{s_{\xi}}$ and the null expansion $\theta_k$
of the section $S_0 \defi \{t=0 \}$ vanish identically.
Using the transformation
Lemma \ref{section} we can write down $s_{\xi}[f]$ and
$\theta_k[f]$ for any surface $S[f]$ defined by a graph
function $f : \mathbb{S}^2 \rightarrow \mathbb{R}$. Substitution
into (\ref{stability2}) (with $\N^{\ell} =0$) gives the   stability
operator $L_{S[f]}$. Explicitly
\begin{eqnarray*}
L_{S[f]} \psi = - \triangle_{\gamma} \psi + 2 W \langle \nabla^{\gamma} f,
\nabla^{\gamma} \psi \rangle  + \left ( W \triangle_{\gamma} f + 2 \langle \nabla^{\gamma}
W, \nabla^{\gamma} f \rangle - W^2 ||\nabla^{\gamma} f||^2 \right ) \psi.
\end{eqnarray*}
We first notice that the principal eigenvalue of the section $\{ f=0 \}$
is $\lambda_{S_0} = 0$. Next,
select $f$ to be invariant  under the same Killing vector of $(\mathbb{S}^2,\gamma)$ 
as $W$. Then $W d f$ is a closed one-form in $\mathbb{S}^2$. Define $V$ as
any solution of $dV = - 2 W df$ and let $U = e^V$. The
stability operator becomes
\begin{eqnarray*}
L_{S[f]} \psi = - e^{-V} \mbox{div}_{\gamma} \left ( e^V \nabla^{\gamma} \psi \right )
+ \left ( W \triangle_{\gamma} f 
+ 2 \langle \nabla^\gamma W, \nabla^\gamma f \rangle -
 W^2 || \nabla^{\gamma} f ||^2 \right ) \psi
\end{eqnarray*}
This is a self-adjoint operator with respect to the $L^2$ product with
measure $\bm{\eta}_V \defi e^V \bm{\eta_{\gamma}}$. It follows that the principal
eigenvalue is given by the Rayleigh-Ritz quotient
\begin{eqnarray}
\label{R-R}
\lambda_{S[f]}  = \inf_{\psi \neq 0}
\frac{\int_{\mathbb{S}^2} || \nabla^{\gamma} \psi ||^2 + 
 \left (  W \triangle_{\gamma} f +  2 \langle \nabla^\gamma W, \nabla^\gamma f \rangle -
 W^2 || \nabla^{\gamma} f ||^2 \right ) \psi^2 \bm{\eta_V}}{\int_{\mathbb{S}^2}
\psi^2  \bm{\eta_V}}.
\end{eqnarray}
To prove the lemma, we
only need to find a pair of functions $\{W,f \}$ for which $\lambda_{S[f]}\neq 0$.
Choose $f = a \cos \theta$ and $W =  b \cos \theta$, where $\{\theta,\varphi\}$
are standard angular coordinates on the sphere and $\{a,b\}$ are
constants. Choose $\psi =1$ in the
Rayleigh-Ritz quotient (\ref{R-R}). This gives an upper
bound for $\lambda_{S[f]}$, namely
\begin{eqnarray*}
\lambda_{S[f]} \leq \frac{\int_{0}^{\pi} 
e^{ c \cos^2 \theta} \left (  2 c \left ( \cos^2 \theta -
\sin^2 \theta \right ) - c^2 \sin^2 \theta  \cos^2 \theta \right )  \sin \theta d \theta}{\int_{0}^{\pi}
e^{c \cos^2 \theta } \sin \theta d \theta } \defi I(c)
\end{eqnarray*}
where $c \defi - a b$.  Expanding the integrands near $c=0$, the following
expression is obtained
\begin{eqnarray*}
I(c) = - \frac{2 c}{3} + 
\frac{2 c^2}{9} + o (c^{3} ).
\end{eqnarray*}
If $c$ is positive and close to zero, then $I(c) <0$ and the corresponding
eigenvalue is also negative, which proves the claim. $\hfill \Box$

\vs

We have shown before that degenerate non-evolving horizons are necessarily
marginally stable. It is natural to ask whether non-degenerate horizons
should be non-marginally stable. The following Proposition states that
non-degenerate, marginally stable, non-evolving horizons can be characterized
by the existence of minimal sections. More precisely,
\begin{theorem}
\label{minimality}
Let $(\H,\ell)$ be a non-evolving horizon satisfying the
topological condition $(\star)$  with constant
surface gravity $\kappa_{\ell} \neq  0$. If $\H$ admits a minimal section (i.e.
a section $S_0$ with vanishing mean curvature vector $H|_{S_0} =0$),
then $\H$ is marginally stable.

Conversely, if $\H$ is marginally stable, then there exists
a section $S[f]$ satisfying 
\begin{eqnarray}
\int_{S[f]} \theta_k[f] \bm{\eta_{S[f]}} = 0. \label{integral}
\end{eqnarray}
If, moreover, for some section 
$S_0$, the Hodge decomposition
$\bm{s_{\ell}} = \bm{z} + d Q_{\ell}$ satisfies (i) $\bm{z} (\nabla^h Q_{\ell} )=0$
and (ii) $\bm{z} ( \nabla^h \phi_0)=0$, where $\phi_0$ is the principal eigenfunction
of  $L_{S_0}$, then the section $S[f]$ is, in fact, minimal.
\end{theorem}

\vs 

{\it Proof.} Assume that $\H$ admits a section $S_0$ which vanishing
mean curvature. In particular, its null expansion $\theta_{k}$ vanishes. The
same proof as for Proposition \ref{marginal} implies $\lambda_{S_0} =0$, and the first claim is proved.

For the (partial) converse, assume that $\lambda_{\H}=0$. 
Choose any section $S_0$  and decompose $ \bm{s_{\ell}} = \frac{du_{\ell}}{2u_{\ell}} + \bm{z}$. According to Lemma \ref{section},
for any section $S[f]$ defined by a graph on $S_0$
 the connection one-form
is $\bm{s_{\ell}[f]} = \pi_f^{\star} \left ( \bm{s_{\ell}} + \kappa_{\ell} d f \right )$.
Its Hodge decomposition is therefore 
$\bm{s_{\ell}[f]} = \frac{d u_{\ell}[f]}{2 u_{\ell}[f]} + \bm{z[f]}$,
where 
\begin{eqnarray}
\label{Transuell}
u_{\ell}[f] = (e^{2 \kappa_{\ell} f} u_{\ell} ) \circ \pi_f, \quad 
\quad \bm{z[f]} = \pi_f^{\star} (\bm{z})
\end{eqnarray}
(this shows in particular
that $\bm{z}$ is independent of the section in the constant surface gravity
case). Let $\phi_0$ a principal eigenfunction of $L_{S_0}$ and define 
\begin{eqnarray*}
f \defi \frac{1}{\kappa_{\ell}} \mbox{ln} \left ( \frac{\phi_0}{u_{\ell}} \right ).
\end{eqnarray*}
Proposition \ref{Dependence_section} gives
\begin{eqnarray*}
L_{S[f]} \left ( u_{\ell}[f] \right ) = 
e^{\kappa_{\ell}f} L_{S_0} \left ( e^{\kappa_{\ell} f} u_{\ell}  \right ) \circ \pi_f = 
e^{\kappa_{\ell}f} L_{S_0} \left ( \phi_0 \right ) \circ \pi_f = 0.
\end{eqnarray*}
Now, the left-hand side can be evaluated using (\ref{LS1}) applied to the section $S[f]$. It follows
\begin{align}
\theta_k[f] & =  
\frac{4}{\kappa_{\ell}} \bm{z[f]} \left ( \nabla^h \mbox{ln} 
(u_{\ell}[f]) \right ) \nonumber \\
& = 
\frac{4}{\kappa_{\ell}} \bm{z}  \left ( \nabla^h \mbox{ln} 
\left (  e^{2 \kappa_{\ell} f} u_{\ell}\right ) \right ) \circ \pi_f \nonumber \\
& =  \frac{4}{\kappa_{\ell}} \bm{z}  \left ( \nabla^h \mbox{ln} 
\left ( \frac{\phi_0^2}{u_{\ell}}  \right ) \right ) \circ \pi_f .
\label{exprethetak}
\end{align}
The integral of the right hand side on $S[f]$ is identically zero.
This proves (\ref{integral}). The last claim is immediate from
(\ref{exprethetak}). \hfill $\Box$

\vs

\noindent {\bf Remark.} This Proposition implies that marginally stable, non-degenerate,
non-evolving horizons can be locally foliated by sections with vanishing total null mean curvature.
Whenever (i) and (ii) are satisfied this foliation is  by minimal sections. This follows
because given a section with these properties a local foliation is 
obtained by dragging along the null normal $\ell$ or, alternatively, by noticing that
the proof applies also to the graph function $f = \kappa_{\ell}^{-1} \mbox{ln}  ( \phi_0 u^{-1} ) + a$
with $a$ an arbitrary constant.

\vs  In expression (\ref{integrate}) we have found an integral equality for $\theta_k$ on any section 
of a non-evolving horizon. This inequality implies, in particular, that on any 
stable section $S_0$ we have $\int_{S_0} \kappa_{\ell} \theta_k \phi_0 {\bm \eta_{S_0}} \leq 0$ and 
zero if and only if the section is marginally stable. This type of inequalities are useful but have
the potential drawback that they involve the principal eigenfunction, which typically is not known
explicitly. Our final aim in this section is to obtain an integral expression which involves computable 
functions on any section $S_0$. 

Let $(\H,\ell)$ be a non-evolving horizon and let $S_0$ any section. Let $\phi_0$ the principal eigenvalue
of $L_{S_0}$. Applying (\ref{LS1}) to $\psi=\phi_0$ we get
\begin{eqnarray*}
- \frac{1}{2} \kappa_{\ell} \theta_k \phi_0 = \lambda_{S_0} \phi_0 + \mbox{div}_h \left ( u_{\ell} \nabla^{h} \left ( 
\frac{\phi_0}{u_{\ell}} \right ) \right ) - 2 {\bm{z}} \left ( \nabla^{h} \phi_0 \right ).
\end{eqnarray*}
In order to get rid of $\phi_0$ we calculate
\begin{align}
- \frac{1}{2} \kappa_{\ell} \theta_k u_{\ell} & = 
\frac{u_{\ell}}{\phi_0} \left ( - \frac{1}{2} \kappa_{\ell} \theta_k \phi_0 \right ) \nonumber \\
 & = \lambda_{S_0} u_{\ell} + \frac{u_{\ell}}{\phi_0} \left [ \mbox{div}_h \left ( u_{\ell} 
\nabla^h \left (
\frac{\phi_0}{u_{\ell}} \right) \right ) - 2 {\bm{z}} \left ( \nabla^h \phi_0 \right )  \right ] \nonumber \\
& = \lambda_{S_0} u_{\ell} + \mbox{div}_h \left ( \frac{u_{\ell}^2}{\phi_0} \nabla^h \left ( \frac{\phi_0}{u_{\ell}} \right ) \right ) 
+ \frac{u_{\ell}^3}{\phi_0} \left \| \nabla^h \left ( \frac{\phi_0}{u_{\ell}} \right ) \right \|_{h}^2 
- \frac{2 u_{\ell}}{\phi_0} \bm{z} \left ( \nabla^h \phi_0 \right ). \label{thetaku}
\end{align}
This identity implies the following result.

\begin{Proposition}
\label{integralthetak}
Let $(\H,\ell)$ be a non-evolving horizon satisfying the topological condition $(\star)$
and $S_0$ a section. Assume that the one-form
$\bm{z}$ in the Hodge decomposition $\bm{s_{\ell}} = \frac{du_{\ell}}{2u_{\ell}} + \bm{z}$
satisfies $\bm{z} ( \nabla^h \phi_0 )=0$, where $\phi_0$ is the principal
eigenfunction of $L_{S_0}$. Then
\begin{itemize}
\item[(i)] If $S_0$ is strictly stable then
$\int_{S_0} \kappa_{\ell} \theta_k u_{\ell} < 0$.
\item[(ii)] If $S_0$ is marginally stable then
$\int_{S_0} \kappa_{\ell} \theta_k u_{\ell} \leq 0$,
and it vanishes if and only if $u_{\ell}$ is a principal eigenfunction of $L_{S_0}$.
\item[(iii)] If $S_0$ is marginally stable and $\kappa_{\ell}$ is constant and non-zero 
on $S_0$, then $\int_{S_0} \kappa_{\ell} \theta_k u_{\ell} \leq 0$,
and zero if and only if $\theta_k=0$.
 
\end{itemize}
\end{Proposition}
{\it Proof.} Integrating (\ref{thetaku}) and using $\bm{z} (\nabla^h \phi_0 )=0$ yields
\begin{eqnarray*}
- \int_{S_0}\kappa_{\ell} \theta_k u_{\ell} = 2 \lambda_{S_0} \int_{S_0} u_{\ell} \bm{\eta_{S_0}} 
+ \int_{S_0} \frac{2u_{\ell}^3}{\phi_0} \left \| \nabla^h \left ( \frac{\phi_0}{u_{\ell}} \right ) \right \|_{h}^2 
\bm{\eta_{S_0}}.
\end{eqnarray*}
If $\lambda_{S_0} > 0$ then the right-hand side is strictly positive, which proves claim (i).
If $\lambda_{S_0}=0$ the right-hand side is non-negative and
zero if and only if $u_{\ell} = c \phi_0$  for some constant $c$.
This proves claim (ii). For the third claim we only need to show that if the integral vanishes then $\theta_k=0$. Since by
point (ii), $u = c \phi_0$, substitution into (\ref{thetaku}) implies $\theta_k =0$.
\hfill $\Box$

Propositions \ref{minimality} and  \ref{integralthetak} require hypotheses involving the
orthogonality between  ${\bm z}_{\ell}$ and the gradient of various functions (the eigenfunction $\phi_0$
and the Hodge dual function $u_{\ell}$ in Proposition (\ref{minimality}) and $\phi_0$ in Proposition (\ref{integralthetak})
The simplest case
where these hypotheses are satisfied involve Killing horizons of static Killing vectors with sufficiently
simple topology. More precisely

\begin{lemma}
Let $\Hxi$ be a Killing horizon in a spacetime $(\M,\gMm)$. Assume that the Killing vector $\xi$ is integrable and that
the Killing horizon satisfies the topological condition $(\star)$ with $S$ simply connected. Then ${\bm{z}}$ vanishes
on any section of $\Hxi$.
\end{lemma}
{\it Proof.} Let $S_0$ be a section $\Hxi$. Since $\xi$ is nowhere zero on $S_0$, the same holds in a sufficiently small 
neighbourhood thereof. In this neighbourhood, the integrability of $\xi$, namely  ${\bm \xi} \wedge d {\bm \xi} =0$ implies the
existence of a one-form ${\bm\beta}$ such that $d {\bm \xi} = {\bm \xi} \wedge {\bm \beta}$
. ${\bm \beta}$ is defined up to addition of $\omega {\bm \xi}$ where $\omega$ is any scalar function. Thus, we can assume without loss of generality that ${\bm \beta} (k ) = 0$. Taking exterior
derivative in $d {\bm \xi} = {\bm \xi} \wedge {\bm \beta}$ yields ${\bm \xi} \wedge d {\bm \beta} = 0$, or
equivalently $d {\bm \beta} = {\bm \xi} \wedge {\bm \gamma}$ for some one-form ${\bm  \gamma}$. Let $\bm{\hat{\beta}}$ be
the pull-back on $S_0$ of ${\bm \beta}$. From the previous considerations we know that $d \bm{\hat{\beta}} = 0$ (recall
that ${\bm \xi}$ is a normal one-form to $S_0$). Now
\begin{eqnarray*}
{\bm s_{\xi}} (X) = - \frac{1}{2} \langle k, \nabla_X \xi \rangle  = - \frac{1}{4} d {\bm \xi} \left ( k,X \right )
= - \frac{1}{4} \left ( {\bm \xi} \wedge {\bm \beta} \right )
\left ( k, X \right ) = \frac{1}{2} {\bm \beta} (X) = \frac{1}{2}
\bm{\hat{\beta}} (X).
\end{eqnarray*}
Thus ${\bm s_{\xi}} = \frac{1}{2} \bm{\hat{\beta}}$ and hence ${\bm s_{\xi}}$ is a closed one-form. The Hodge decomposition 
${\bm s_{\xi}} = d Q_{\xi} + {\bm z}$ implies that $d {\bm z} =0$. Since ${\bm z}$ is also divergence-free, it follows
that ${\bm z}$ is harmonic. If $S_0$ admits no non-trivial
harmonic one-forms (in particular if $S_0$ is simply connected and compact) then ${\bm z}$ vanishes identically.
\hfill $\Box$

\vs

The second most relevant case where hypothesis (i) and (ii) in Proposition
\ref{minimality} are satisfied involve axially symmetric non-evolving horizons in four spacetime
dimensions and spherical topology. We devote next section to analyze this case and to relate the 
stability properties of such horizons with the area-angular momentum inequalities.

\vs

\section{Axially symmetric MOTS and angular momentum}
\label{Axially}

We start with the following definition, which is essentially the same as the one given in 
\cite{JaramilloReirisDain}.
\begin{definition}
\label{axialMOTS}
A MOTS $(S,h)$ is axially symmetric if there
exists a vector field $\eta \in \mathfrak{X} (S)$ with closed orbits
satisfying
\begin{itemize}
\item[(i)] ${\cal L}_{\eta\, }  h  =0$. 
\item[(ii)] ${\cal L}_{\eta \, } {\bm{s_{\hat{\ell}}}} =0$, for some
choice of basis $\{\hat{\ell},\hat{k} \}$.
\item[(iii)] $\eta$ commutes with the stability operator $L_v$ for
some choice of normal vector $v$ of the form (\ref{vstar}).
\end{itemize}
\end{definition}

\vs 

\begin{definition}
The {\bf angular momentum} of an axially symmetric, two-dimensional
MOTS $S$ is the integral
\begin{eqnarray}
J(S) = \frac{1}{8 \pi} \int_S {\bm{s_{\ell}}} (\eta) \bm{\eta}_{S},
\label{J}
\end{eqnarray}
where the connection one-form  $\bm{s_{\ell}}$
is defined with respect to any basis $\{ \ell,k \}$.
\end{definition}

\vs 

\noindent {\bf Remark.} 
The independence of the angular momentum with respect to the
choice of basis can be found, e.g. in \cite{Jaramillo2011}. The
argument uses only the divergence-free character of $\eta$ and can be 
described as follows.  Using $\bm{s}_{\ell} = d Q_{\ell} + \bm{z}$ yields
\begin{align}
J(S) & = \frac{1}{8\pi} \int_S \bm{s_{\ell}} (\eta) \bm{\eta}_{S} =
\frac{1}{8 \pi}
\int_S \left ( d Q_{\ell} (\eta) + \bm{z} (\eta ) \right ) \bm{\eta}_{S} =
\frac{1}{8 \pi}
\int_S \left ( \mbox{div}_h \left ( \eta Q_{\ell} \right )
+ \bm{z} (\eta ) \right ) \bm{\eta}_{S} = \nonumber \\
& = \frac{1}{8 \pi}\int_S \bm{z} (\eta) \bm{\eta}_{S}.
\label{Jalternative} 
\end{align}
The last term only involves ${\bm z}$, which is independent
of the choice of basis $\{\ell,k\}$.

%

\vs The following theorem establishes a remarkable inequality between
area and angular momentum for stable, axially symmetric  MOTS in four
dimensional spacetimes.

\begin{theorem}[Jaramillo, Reiris, Dain, \cite{JaramilloReirisDain}]
\label{Pepe}
Let $(M,\gM)$ be a spacetime satisfying the dominant energy condition.
Let $S$ a two-dimensional MOTS in $(M,\gM)$ and
assume $S$ to be an axially symmetric and stable with
respect to the null direction $- \frac{1}{2} \hat{k}$. Then 
\begin{eqnarray*}
\bm{|S| \geq 8 \pi |J(S)|}.
\end{eqnarray*}
where $|S|$ is the area of $S$.
\end{theorem}

A related statement for stable minimal surfaces embedded in maximal spacelike
hypersurfaces in vacuum spacetimes with non-negative cosmological constant
had been previously obtained by Dain and Reiris \cite{DainReiris}. Previous to those
results,  area-angular momentum inequalities were obtained in the context of stationary and axially 
symmetric black holes by Hennig, Ansorg and Cederbaum \cite{Hennig2008}.
A interesting recent review on inequalities involving area, angular momentum
and mass can be found in \cite{DainReview} (see also \cite{JaramilloReview}).

The following lemma has been used many times in the literature, see
e.g. \cite{JaramilloReirisDain}. We include its
proof for completeness.
\begin{lemma}
Let $S_0$ be an axially symmetric MOTS in a spacetime $(M,\gMm)$
and chose a normal basis $\{ \hat{\ell},\hat{k}\}$ satisfying (ii) in Definition 
\ref{axialMOTS}. Then the
eigenfunction $\phi_v$ of the stability operator $L_v$, with $v$
as in condition (iii) of Definition \ref{axialMOTS}, satisfies ${\mathcal L}_{\eta} \phi_v =0$. 

Moreover, if $S$ is two-dimensional and of spherical
topology, then both $\phi_v$ and the function $Q_{\hat{\ell}}$ in the Hodge
decomposition $\bm{s_{\hat{\ell}}} = d Q_{\hat{\ell}} + \bm{z}$ satisfy
$\bm{z} (\nabla^h \phi_v) = \bm{z} (\nabla^h  Q_{\hat{\ell}} ) =0$.
\end{lemma}
{\it Proof.} Since $L_v$ and ${\mathcal L}_{\eta}$ commute,
${\mathcal L}_{\eta} \phi_v $ is an eigenfunction with principal eigenvalue
$\lambda_v$. Hence, there exists a constant $c$ such that 
${\mathcal L}_{\eta} \phi_v = c \phi_v$.
Integrating on $S$ it follows
\begin{eqnarray}
c \int_S \phi_v \bm{\eta_S} = \int_S {\mathcal L}_{\eta} (\phi_v) \bm{\eta_S} =
\int_S \mbox{div}_h \left ( \phi_s \eta \right ) \bm{\eta_S} = 0.
\label{c=0}
\end{eqnarray}
Thus $c=0$ and ${\mathcal L}_{\eta} \phi_v = 0$. 
For the second part, from ${\mathcal L}_{\eta} {\bm s_{\hat{\ell}}} = 0$
it follows ${\mathcal L}_{\eta} Q_{\hat{\ell}} =0$ because  
\begin{eqnarray}
0 = {\mathcal L}_{\eta} (\mbox{div}_h \bm{s_{\hat{\ell}}}) =
{\mathcal L}_{\eta} (\triangle_h Q_{\hat{\ell}}) = \triangle_h 
\left ( {\mathcal L}_{\eta} Q_{\hat{\ell}} \right )  
\quad \Longrightarrow \quad 
{\mathcal L}_{\eta} Q_{\hat{\ell}} = \mbox{const}
\quad \Longrightarrow \quad 
{\mathcal L}_{\eta} Q_{\hat{\ell}} = 0,
\end{eqnarray}
where the last implication follows by integration on $S$
as in (\ref{c=0}). As a consequence
we also have ${\mathcal L}_{\eta} {\bm z} = 0$. 
If, moreover, $S$  is of spherical topology, then $\bm{z}$ is the Hodge dual of the gradient of a function $W$
and a similar argument as before implies ${\mathcal L}_{\eta} W =0$.
or, equivalently,
that $d W$ is orthogonal to $\eta$.
$S$ being two-dimensional, the Hodge dual $\bm{z}$ of $d W$ is
tangent to $\eta$. 
The statements are now obvious because $\phi_v$ and $Q_{\hat{\ell}}$
are constant along $\eta$. \hfill $\Box$

\vs

All sections in a totally geodesic null horizon $\H$ are isometric. Moreover, the 
transformation law (\ref{first}) implies that $\bm{z[f]}$ is independent of the section.
At first it may seem that this requires $\kappa_{\ell}$ to be constant, but this is not
the case because one can always choose an affinely parametrized null normal $\ell_0$
of $\H$. For this choice, the surface gravity is identically zero and hence (\ref{first}) implies that 
${\bm s_{\ell_0}}$ is independent of the section. Since, moreover, ${\bm z}$ is independent of the choice of
null normal basis $\{\ell,k\}$, the independence of ${\bm z}$ with the section follows. Consequently, 
if $\H$ is three-dimensional and admits sections which are axially symmetric MOTS, then both the
area $|S|$ and the angular momentum $J(S)$ are independent of the choice of axially symmetric
section in $\H$. In this setting we may simply write
\begin{eqnarray}
|S_{\H}| \geq 8 \pi J(\H)
\label{area-angular}
\end{eqnarray}
to refer to the area-angular momentum inequality for any of the axially symmetric section of $\H$.

Since non-evolving horizons have a preferred choice of
null normal $\ell$,  the following definition becomes natural.

\begin{definition}
Let $(\H,\ell)$ be a non-evolving horizon satisfying the topological condition $(\star)$.
$(\H,\ell)$ is called {\bf axially
symmetric} if there exist a section $S_0$ of $\H$ which is an axially
symmetric MOTS and point (ii) in Definition (\ref{axialMOTS}) is satisfied
by the basis $\{ \hat{\ell} , \hat{k} \} = 
\{ \ell|_{S_0}, k|_{S_0} \}$.
\end{definition}

\vs

We can now state and proof the following result, which gives sufficient conditions
on a non-evolving horizon for the validity of (\ref{area-angular}). Its proof
will be the basis for our subsequent clarification of the relationship 
between the argument in Hennig {\it et al.} and the argument
in Jaramillo {\it et al} of their corresponding area-angular momentum inequalities.
 
\begin{theorem}
\label{inequality}
Let $(\H,\ell)$ be an axially symmetric, non-evolving horizon in a four dimensional 
spacetime $(M,g^{(4)})$ satisfying the dominant energy condition
(in particular $(\H,\ell)$ is an isolated horizon). 
Assume
that $\H$ is topologically $\mathbb{S}^{2} \times \mathbb{R}$ with 
$\ell$ tangent to the $\mathbb{R}$ factor and
that the surface gravity $\kappa_{\ell}$ is constant and non-zero.
Then, if  $\int_{S_0}  \kappa_{\ell} \theta_k u_{\ell} {\bm \eta_{s_0}} \leq 0$ for any
section $S_0$, then
\begin{eqnarray}
|S_{\H}| \geq 8 \pi J(\H),
\label{ineqHorizon}
\end{eqnarray}
and equality occurs only if and only if the following four conditions
are satisfied: 
\begin{itemize}
\item[(i)] The metric $h$ of any section
of the horizon reads (in appropriate coordinates)
\begin{equation}
h =  |J| \left ( 1 + \cos^2 \theta \right ) d\theta^2 +
\frac{4 |J| \sin^2 \theta}{1 + \cos^2 \theta} d\varphi^2.
\label{metrich}
\end{equation}
where $J$ is an arbitrary non-zero constant.
\item[(ii)] There exists a section $S_1$ where $\Ein(\ell,k)\eqSone 0$.
\item[(iii)] The normal connection one-form of $S_1$ reads
\begin{eqnarray}
\label{sell}
s_{\ell}= - \frac{\cos \theta \sin \theta}{1 + \cos^2 \theta} d \theta
+ \frac{2 J \sin^2 \theta}{|J| \left ( 1+ \cos^2 \theta \right )^2}
d \varphi.
\end{eqnarray}
\item[(iv)] $\int_{S_1} \theta_k[S_1] \left ( 1 + \cos^2 \theta \right ) 
\bm{\eta_{S_1}} = 0$.
\end{itemize}
\end{theorem}

{\it Proof.}
Let $S_0$ be an axially symmetric section of $\H$. 
Let us start by calculating
$u_{\ell} \| {\bm s_{\ell}} \|^2 + u_{\ell} \, \mbox{div}_h {\bm s_{\ell}}$:
\begin{eqnarray*}
u_{\ell} \| {\bm s_{\ell}} \|_h^2 + u_{\ell} \, \mbox{div}_h {\bm s_{\ell}} =
u_{\ell} \| {\bm s_{\ell}} \|_h^2 + \mbox{div}_h \left ( u_{\ell} {\bm s_{\ell}} \right )
- \bm{s_{\ell}} \left ( \nabla^h u_{\ell} \right ) 
= - \frac{\| \nabla^h u_{\ell} \|_h^2}{4 u_{\ell}} + u_{\ell} \| {\bm z} \|^2_h 
+ \mbox{div}_h \left ( u_{\ell} \bm{s_{\ell}} \right ),
\end{eqnarray*}
where in the last equality we have used ${\bm s_{\ell}}  = \frac{d
u_{\ell}}{2 u_{\ell}} + \bm{z}$ and the orthogonality between $\bm{z}$ and
$d u_{\ell}$.  Inserting this in  expression (\ref{Einxik})
(with $\N^{\ell} =0$) and integrating on $S_0$ it follows
\begin{eqnarray}
\int_{S_0} \left ( \frac{\| \nabla^h u_{\ell} \|_h^2}{2 u_{\ell}} + u_{\ell} \Scal_h  
\right ) {\bm{\eta_{S_0}}} = 
\int_{S_0} \left ( u_{\ell} \Ein(\ell,k) - \kappa_{\ell} \theta_k u_{\ell}
+ 2 u_{\ell} \| {\bm z} \|^2_h \right ) {\bm \eta_{S_0}}.
\label{EqHorizon}
\end{eqnarray}
Under the conditions of the theorem, the first two terms in the right-hand side
are non-negative. 
Discarding them yields
\begin{eqnarray}
  \int_{S_0} \left ( \frac{\| \nabla^h u_{\ell} \|_h^2}{2 u_{\ell}} + u_{\ell} \Scal_h 
\right ) {\bm{\eta_{S_0}}} \geq  
\int_{S_0} 2 u_{\ell} \| {\bm z} \|^2_h  {\bm \eta_{S_0}}.
\label{IneqHorizon}
\end{eqnarray}
The area-angular momentum inequality 
is proved in \cite{JaramilloReirisDain}
by choosing a coordinate system on $S_0$ 
in which  the metric $h$ reads  (this form of the metric was
introduced in \cite{Ash3}, a detailed proof of existence of the coordinate
system appears in \cite{DainReiris}) 
\begin{eqnarray}
h = e^{\sigma} \left ( e^{2q} d \theta^2 + \sin^2 \theta d \varphi^2 \right ),
\label{generalh}
\end{eqnarray}
where $\sigma, q$ are functions of $\theta$ satisfying $q + \sigma = c$, where
$c$ is constant (see expression (13) in \cite{JaramilloReirisDain}). The
inequality $|S_0| \geq 8 \pi J(S_0)$ is proved in that paper
as a consequence of the inequality
\begin{eqnarray}
\int_{S_0} \left ( \| \nabla^h \alpha \|^2_h + \frac{\alpha^2}{2} \Scal_h \right ) \bm{\eta_{S_0}}
\geq \int_{S_0} \alpha^2 \| z \|^2 \bm{\eta_{S_0}} \label{Dain}
\end{eqnarray}
where $\alpha$ is an arbitrary function which is then chosen to be related to the
metric $h$ by $\alpha = e^c e^{-\sigma/2}$.
This inequality, in turn, follows from
the stability of the MOTS along $-\frac{1}{2} \hat{k}$.
The close relationship between (\ref{IneqHorizon}) and (\ref{Dain})
is apparent.  The freedom in $u_{\ell}$ in (\ref{IneqHorizon}) comes
from the freedom in choosing the section $S_0$. Given the transformation law (\ref{Transuell})
for $u_{\ell}$, we can choose the graph function $f$ so that
\begin{equation}
u_{\ell} [f] = u_{\ell} e^{2\kappa_{\ell} f} = \alpha^2 = e^{2c} e^{- \sigma}. \label{uf}
\end{equation}
The section $S_1 \defi S[f]$ is still axially symmetric.
The argument in \cite{JaramilloReirisDain} applied to $S_1$ proves
$|S_1| \geq 8 \pi J(S_1)$. Since the inequality is independent of the
(axially symmetric) section, (\ref{ineqHorizon}) follows.

For the equality case, it is proved
in \cite{JaramilloReirisDain}, \cite{Acena} that 
equality in (\ref{Dain})
occurs if and only if the following two conditions hold:
(i) the geometry on $S_1$ (and hence
of any section of the horizon) is isometric to the extreme Kerr throat 
geometry, given explicitly by (\ref{metrich}) and (ii)
the one-from $\bm{z}$  takes the form
\begin{eqnarray}
\bm{z} = \frac{2 J \sin^2 \theta}{|J| \left ( 1 + \cos^2 \theta
\right )^2} d \varphi.
\label{z}
\end{eqnarray}
In our setting we have, in addition, discarded
two non-negative terms in (\ref{IneqHorizon}). This forces
$\Ein (\ell,k) \eqSone 0$ and 
$\int_{S_1} \kappa_{\ell} \theta_k[S_1] u_{\ell}[S_1] {\bm \eta_{S_1}}= 0$. Moreover, 
since $u_{\ell}[f] = e^{2c} e^{-\sigma}$  it follows from (\ref{metrich})
and  (\ref{generalh}) that
\begin{equation*}
u_{\ell}[S_1] = |J| \left ( 1 + \cos^2 \theta \right ).
\end{equation*}
Inserting this function and (\ref{z}) into $s_{\ell} =
\frac{d u_{\ell}[S_1]}{2 u_{\ell}[S_1]} + \bm{z}$ gives (\ref{sell}). 
Condition (iv) follows directly from  
$\int_{S_1} \kappa_{\ell} \theta_k[S_1] u_{\ell}[S_1] {\bm \eta_{S_1}}= 0$
and the explicit form of $u_{\ell}$.
\hfill $\Box$

\vs 
\noindent {\bf Remark.} As mentioned above, inequality (\ref{Dain}) is proven 
in \cite{JaramilloReirisDain} from the stability of the MOTS by using
direct estimates. An alternative derivation can be obtained from
the Rayleigh-Ritz type characterization of the principal 
eigenvalue obtained in expression (16) in \cite{AMS2008}
and using the fact that, whenever the function $u$ in that paper is
axially symmetric then $\omega[u]  = 0$.

\vs

Combining this theorem with Proposition \ref{integralthetak} yields the following Corollary.

\begin{corollary}
\label{inequalitystable}
Let $(\H,\ell)$ be an axially symmetric, non-evolving horizon in a four dimensional 
spacetime $(M,g^{(4)})$ satisfying the dominant energy condition. 
Assume
that $\H$ is topologically $\mathbb{S}^{2} \times \mathbb{R}$ with
$\ell$ tangent to the $\mathbb{R}$ factor and
that the surface gravity $\kappa_{\ell}$ is constant and non-zero.
If $\H$ is stable then
\begin{eqnarray*}
|S_{\H}| \geq 8 \pi J(\H),
\end{eqnarray*}
and equality occurs if and only if the following condition hold
\begin{itemize}
\item[(i)] The horizon is marginally stable.
\item[(ii)] The metric $h$  of any section
of the horizon reads as in (\ref{metrich}).
\item[(iii)] There exists a minimal section $S_1$ (i.e. a section satisfying $\theta_k[S_1] =0$).
\item[(iv)] The Einstein tensor satisfies $\Ein (\ell,k) \eqSone 0$.
\item[(v)] The normal connection one-form of $S_1$ reads as in (\ref{sell}).
\end{itemize}
\end{corollary}

We are now in a position where we can explain in which sense Theorem \ref{inequality} clarifies the relationship
between the area-angular momentum inequality of Hennig, Ansorg and Cederbaum \cite{Hennig2008} and
the area-angular momentum inequality obtained by Jaramillo, Reiris and Dain \cite{JaramilloReirisDain}.

The setup in \cite{Hennig2008} deals with stationary and axially symmetric black hole
spacetimes which are vacuum in a neighbourhood of the event horizon. The whole argument is performed in adapted coordinates
where the metric takes the following form.
\begin{align*}
ds^2 & = \left ( \frac{a}{\hat{u}} + \hat{u} b^2 \sin^2 \theta \right ) dR^2 - \frac{\Delta}{\hat{u}} d\tilde{t}^2 + \hat{u} 
\sin^2 \theta \left ( d \tilde{\varphi} - \omega d \tilde{t} \right )^2 + \hat{\mu} d\theta^2 \\
& + 2 \left ( \frac{T}{\hat{u}} + \omega \hat{u} b \sin^2 \theta \right ) dR d \tilde{t} - 2 \hat{u} b \sin^2 \theta
dR d \tilde{\varphi}
\end{align*}
where $\Delta = 4(R^2 - r_h^2)$, $r_h$ is a positive constant which  defines the location of the horizon $\H$ (at $R=r_h$).
The functions $a,b, \hat{u}, \hat{\mu}, \omega$ depend on $\{R,\theta\}$ and satisfy $\hat{u}>0$, $\hat{\mu}>0$,
$\omega |_{\H} = \omega_{h}$ constant
and $\frac{2 r_h}{\sqrt{\hat{\mu}\hat{u}}} = \kappa> 0 $ constant. In fact, $\kappa$ is precisely the surface gravity of the Killing
vector for which $\H$ is a Killing horizon, namely $\xi = \partial_{\tilde{t}} + \omega_h \partial_{\tilde{\varphi}}$.  The function $T$
depends on $R$ alone and satisfies $T(R=r_h)= \frac{4 r_h}{\kappa}$ and $\left . \frac{dT}{dR} \right |_{R=r_h} < 0$.
The sections
of the horizon considered in that paper are ${\cal S} \defi \{ \tilde{t} = \mbox{const.}\} $. The key condition
imposed  by Hennig {\it at al} is the {\it subextremality} of the horizon, namely the existence of inward variations from 
${\cal S}$ which strictly decrease the null expansion $\theta_k$ at every point (this is known to be
equivalent to the strict stability of the horizon, see Proposition 5.1 in \cite{AMS2008}). In fact, the authors do not quite
need this condition, but a weaker condition stated in Lemma 3.1 in \cite{Hennig2008}, namely 
\begin{equation}
\int_0^\pi \left . \frac{\partial \left ( \hat{\mu} \hat{u} \right )}{\partial R} \right |_{R=r_h} \sin \theta d \theta > 0.
\label{integralcondition}
\end{equation}
This inequality is sufficient to prove $|{\cal S}| > 8 \pi J(S)$ \cite{Hennig2008}. Now, we can understand 
this result in the light of Theorem \ref{inequality}. It is matter of straightforward calculation to determine  the induced metric $h$,
normal connection one-form  ${\bm s_{\xi}[{\cal S}]}$ and null expansion $\theta_k[{\cal S}] $ of ${\cal S}$.
Letting $\tilde{u} \defi \hat{u} |_{r=r_h}$, the result is 
\begin{align*}
h & = \frac{4 r_h^2}{\kappa^2 \tilde{u}} d \theta^2 + \tilde{u} \sin^2 \theta d \tilde{\varphi}^2, \\
{\bm s_{\xi}[{\cal S}]} & = - \frac{1}{2\tilde{u}} \frac{\partial \tilde{u}}{\partial \theta} d \theta 
+ \frac{\kappa \tilde{u}^2 \sin^2 \theta}{8 r_h} \left . \frac{\partial \omega}{\partial R} \right |_{R=r_h} d \tilde{\varphi}, \\
\theta_k[{\cal S}] & = - \frac{\kappa^3 \tilde{u}}{16 r_h^3} \left . \frac{\partial \left ( \hat{u} \hat{\mu} \right )}{\partial R} 
\right |_{R=r_h}.
\end{align*}
According to the definition of $u_{\ell}$ we conclude that $u_{\ell} [{\cal S}] = a_0 \tilde{u}^{-1}$  where $a_0$ is an arbitrary
positive constant. We observe, first of all, that the metric $h$ has the form  (\ref{generalh}) with $e^{\sigma} = \tilde{u}$,
$e^{q} = \frac{2r_h}{\kappa \tilde{u}}$ and $e^c = \frac{2r_h}{\kappa}$. Moreover, condition (\ref{uf}) is automatically satisfied
if we choose $a_0 = \frac{4 r_h^2}{\kappa^2}$. Consequently, we have
\begin{equation}
-\int_{\cal S}  \kappa_{\ell} u_{\ell}[{\cal S}] \theta_k [{\cal S}] \bm{\eta_{\cal S}} = 
\int_{\cal S} \frac{\kappa^2}{4 r_h} \left . 
\frac{\partial \left ( \hat{u} \hat{\mu} \right )}{\partial R} \right |_{R=r_h} \sin \theta d \theta 
d \tilde{\varphi} = \frac{\pi \kappa^2}{2r_h} \int_0^{\pi} 
\left . \frac{\partial \left ( \hat{u} \hat{\mu} \right )}{\partial R} 
\right |_{R=r_h} \sin \theta d \theta 
\end{equation}
so that the integral condition (\ref{integralcondition}) is precisely the same as the requirement 
$\int_{{\cal S}} \kappa_{\ell} \theta_k[{\cal S}] u_{\ell} [{\cal S}] {\bm \eta_{{\cal S}}} < 0$, which is
directly related to the main hypothesis of Theorem \ref{inequality}. As we have seen
along the proof of this theorem, the key inequality behind $|S| \geq 8 \pi J(S)$
is (\ref{Dain}). In one case, this inequality follows from the stability of the MOTS and
a suitable choice of coordinates in $S$ and a choice of $\alpha$
expressed in terms of the metric.
In the other case,
the inequality follows form $\int_{S_0} \kappa_{\ell} \theta_k u_{\ell} {\bm \eta_{S_0}} \leq 0$
after exploiting the freedom in choosing the section of the horizon.
It is remarkable that the coordinate system
used in \cite{Hennig2008} has the property that the sections $\tilde{t} = \mbox{const.}$ are precisely
the sections $S[f]$ for which the integral condition  - $\int_{S[f]} \kappa u_{\ell}[f] \theta_k[f] \bm{\eta_{S[f]}} > 0$
becomes exactly the integral in the left-hand side of (\ref{Dain}).

As mentioned in the Introduction, the relationship between the proof in \cite{Hennig2008} and the proof
of the area-angular momentum inequality for stable minimal surfaces lying in maximal vacuum initial data
sets \cite{DainReiris} has been clarified recently in \cite{ChruscielEckstein} by showing 
in an appropriate coordinate system that inequality (\ref{integralcondition}) is precisely
the strict version
of (\ref{Dain}). The latter is, in this setting,  a consequence of the Rayleigh-Ritz quotient
characterization of the principal eigenvalue of the stability operator for minimal surfaces. The clarification we have 
obtained
in this paper provides, in addition, a clear geometric interpretation of (\ref{integralcondition})
in terms of the geometry of the Killing horizon.

A final remark is in order. In \cite{ChruscielEckstein} the final step for the existence of singularities in 
two-Kerr black hole spacetime has been achieved. In \cite{HennigNeugebauer2009}, \cite{HennigNeugebauer2011} and \cite{HennigNeugebauer2012}
it was proved that the double Kerr solution of Kramer and Neugebauer \cite{KramerNeugebauer1980} is the only possible candidate
for a stationary and axially symmetric  asymptotically flat black hole with an event horizon of two connected components. 
Moreover, the authors also prove that either if both components are degenerate or if the strict inequality 
$|S| > 8 \pi J(S)$ holds on each non-degenerate component, then there must exist a conical singularity
in the portion of the axis of symmetry lying between each connected component of the event horizon. The proof
was based on explicit formulae obtained previously by Manko and Ruiz \cite{MankoRuiz2011} where existence
of singularities was shown under positivity of the Komar masses of each
black hole constituent. In \cite{ChruscielEckstein} the stability (in the sense
of MOTS) of each connected component
of the event horizon is proved as a consequence of existence
of an outermost MOTS in spacelike asymptotically flat slices
\cite{AnderssonMetzger, Eichmair, AnderssonMetzgerEichmair}.
Stability only implies $|S| \geq 8 \pi J(S)$ so one might think that
the results by Hennig and Neugebauer are, by themselves, not quite
sufficient to finish the proof. In \cite{ChruscielEckstein} this issue is dealt with 
by mentioning (with no explicit proof) that the arguments in by
Hennig and Neugebauer can be extended to cover the case 
$|S| = 8 \pi J(S)$ on non-degenerate components.

In view of the results in this paper, an alternative argument to exclude the
equality case $|S| = 8\pi J(S)$ in non-degenerate components is to show
that any of the five conditions (i)-(v) in Corollary \ref{inequalitystable}
is not satisfied in the double Kerr spacetime.
The coordinate transformation $\cos \theta = a_0 (\zeta- \zeta_0)$, with $a_0,\zeta_0$ constants brings
the metric (\ref{metrich}) into the form
\begin{equation}
\label{form}
h= \frac{a_1^2}{F(\zeta)}  d\zeta^2 + a_2^2 F(\zeta)  d\varphi^2, 
\end{equation}
with $a_1 \defi a_0 \sqrt{|J|}$, $a_2 \defi 2 \sqrt{|J|}$, and
\begin{eqnarray*}
F(\zeta) \defi \frac{1 - a^2_0 \left ( \zeta - \zeta_0 \right )^2}{1+ a_0^2 \left ( \zeta- \zeta_0 \right )^2}. 
\end{eqnarray*}
In fact, it is straightforward to check that $\cos \theta = a_0 (\zeta- \zeta_0)$
 is the most general transformation that brings (\ref{metrich}) into the form 
(\ref{form}). On the other hand the geometry of a non-degenerate 
connected component of the ``event horizon'' 
of the double Kerr solution is 
\begin{align}
\tilde{h} = \frac{a_1^2}{\tilde{F}(\zeta)}  d \zeta^2  + a_2^2 \tilde{F} (\zeta) d\varphi^2,
\label{induced}
\end{align}
where $-\tilde{F}(\zeta)$ is the real part of the Ernst potential associated to the
asymptotically timelike Killing vector $\hat{\xi}$. Expression (\ref{induced}) follows by setting
$\{ \rho = 0, t=0\}$ in the spacetime metric written
in Weyl-Papapetrou coordinates  $\{t,\varphi,\rho,\zeta\}$  as written e.g. in formula (1) in \cite{HennigNeugebauer2012} and
using the facts that the functions $k$ and $a$ in that metric are constant along the horizon. Now, the
Ernst potential associated to the Killing vector $\hat{\xi}$ for the double Kerr solution is well-known.
Its explicit form is given, for instance, in expression (34.94)  in
\cite{ExactSolution}. Restricting to 
$\rho=0$, $\zeta \in (K_4, K_3)$ ($K_4 < K_3 < K_2 \leq K_1$ are constants)
which corresponds to a non-degenerate
component of the horizon, and taking the real part yields 
\begin{equation*}
\tilde{F} (\zeta) = \frac{(K_1 - \zeta)(K_2 - \zeta)
( K_3 -  \zeta ) (\zeta - K_4 )}{P_4 (\zeta)},  
\end{equation*}  
where $P_4(\zeta)$ is a fourth-order polynomial which does not
vanish anywhere in the interval $\zeta \in (K_4,K_3)$.
It is obvious that $\tilde{h}$ is not isometric to $h$ in (\ref{form}), so we are not in the equality case of the
area-angular momentum inequality, and non-degenerate horizons in the double Kerr spacetime necessarily
satisfy the strict inequality $|S| > 8 \pi J(S)$.

\section{Acknowledgments}

I am very grateful to Jos\'e Luis Jaramillo for very interesting
and useful comments on the manuscript and for provinding several
references. Financial support under the
projects FIS2009-07238 (Spanish MICINN)
and P09-FQM-4496 (Junta de Andaluc\'{\i}a and FEDER funds) are gratefully
acknowledged.


\end{document}